\documentclass[usenames]{lmcs}
\pdfoutput=1

\usepackage{lastpage}
\renewcommand{\dOi}{14(3:7)2018}
\lmcsheading{}{1--\pageref{LastPage}}{}{}%
{Dec.~12,~2017}{Aug.~03, 2018}{}

\usepackage[all]{xy}
\usepackage{mathpartir}
\usepackage{amsmath}
\usepackage{ulem}
\usepackage{graphics}
\usepackage{color}
\usepackage{amssymb}
\usepackage{prooftree}
\usepackage{float}

\newcommand{\cal}[1]{\mathcal #1}

\long\def\ignore#1{\relax}

\newcommand{\s}{{\tt t}}
\newcommand{\uu}{{\tt u}}
\newcommand{\vv}{{\tt v}}
\newcommand{\x}{{\tt x}}
\newcommand{\y}{{\tt y}}
\newcommand{\z}{{\tt z}}
\newcommand{\w}{{\tt w}}
\newcommand{\ap}{{\tt a}}
\newcommand{\bp}{{\tt b}}
\newcommand{\cp}{{\tt c}}
\newcommand{\iid}{{\tt I}}
\newcommand{\dup}{{\tt D}_{\tt up}}
\newcommand{\Ap}{{\mathcal A}}

\newcommand{\Rew}[1]{\rightarrow_{#1}}

\newcommand{\nf}{{\tt nf}}
\newcommand{\hnf}{{\tt hnf}}

\newcommand{\redb}{\rightarrow_{\beta}}
\newcommand{\redo}{\rightarrow_{\beta\omega}}

\newcommand{\redbs}{\rightarrow^{*}_{\beta}}

\newcommand{\A}{{\tt A}}
\newcommand{\D}{{\tt B}}
\newcommand{\arrow}{\rightarrow}
\newcommand{\inter}{\wedge}

\newcommand{\iset}[1]{\{#1\}}
\newcommand{\der}{\vdash}
\newcommand{\ders}{\vdash_{{\footnotesize \Su}}}

\newcommand{\derw}{\vdash_{\footnotesize w}}

\newcommand{\sep}{\hspace{.5cm}}
\newcommand{\Mu}{\mathcal H}

\newcommand{\Su}{\mathcal S}

\newcommand{\dem}{\triangleright}
\newcommand{\demnf}{\triangleright_{\tt nf}}

\newcommand{\multiset}[1]{[#1]}

\newcommand{\emul}{\multiset{\, }}
\newcommand{\set}[1]{\{ #1 \}}
\newcommand{\ie}{{\it i.e.}}
\newcommand{\cf}{{\it cf.}}
\newcommand{\eg}{{\it e.g.}}
\newcommand{\ih}{{\it i.h.}}
\newcommand{\isubs}[2]{\{ #1 / #2 \}}

\newcommand{\Gam}{\Gamma}

\newcommand{\Del}{\Delta}
\newcommand{\del}{\del}

\newcommand{\fv}[1]{{\tt fv}(#1)}

\newcommand{\dom}[1]{{\tt dom}(#1)}

\newcommand{\pathd}[1]{\times_{2}}

\newcommand{\sm}{\setminus}

\newcommand{\es}{\emptyset}

\def\l{\lambda}

\newcommand{\sig}{\sigma}

\newcommand{\ccontext}{{\tt C}}

\newcommand{\ccount}[1]{\#(#1)} 
\newcommand{\iI}{i \in I}

\newcommand{\bu}{\bigvee}

\newcommand{\Head}{{\small {\tt Head}}}
\newcommand{\StartHead}{{\small {\tt Head}}}

\newcommand{\Abs}{{\small {\tt Abs}}}
\newcommand{\Union}{{\small {\tt Union}}}

\newcommand{\K}{{\tt T}}

\newcommand{\KI}{{\tt TI}}
\newcommand{\LH}[2]{{\tt H}^{#1}{#2}}

\newcommand{\meas}{{\tt meas}}

\newcommand{\occ}[1]{\mathtt{o}(#1)}
\newcommand{\tocc}[1]{\mathtt{to}(#1)}

\newcommand{\ems}{\multiset{\;}}
\newcommand{\axvar}{\mathtt{var}}
\newcommand{\axvarw}{\mathtt{var}_w}
\newcommand{\arri}{\mathtt{\arrow_{I}}}
\newcommand{\arrie}{\mathtt{\arrow_{I\multiset{}}}}
\newcommand{\arrine}{\mathtt{\arrow_{I\not\multiset{}}}}
\newcommand{\arre}{\mathtt{\arrow_{E}}}
\newcommand{\arree}{\mathtt{\arrow_{E\emm}}}
\newcommand{\arrene}{\mathtt{\arrow_{E\not\emm}}}
\newcommand{\inti}{\mathtt{\wedge_{I}}}
\newcommand{\inte}{\mathtt{\wedge_{E}}}
\newcommand{\many}{\mathtt{m}}

\newcommand{\mult}[1]{[#1]}
\newcommand{\tyj}[3]{{#2} \vdash #1{:}#3}
\newcommand{\emm}{ [\,] }
\newcommand{\real}{\Vdash}
\newcommand{\interp}[1]{[ \! [ #1 ] \! ]}
\newcommand{\Hn}[4]{{\tt H}^{#1:\multiset{#2 #3}}(#4,#3)}
\newcommand{\Hnp}[4]{{\tt H}^{#1:\multiset{#2}}(#3,#4)}

\title{Inhabitation for Non-idempotent Intersection Types}

\author[Antonio Bucciarelli]{Antonio Bucciarelli{\rsuper{a}}}

\author[Delia Kesner]{Delia Kesner\rsuper{b}}	
\address{{\lsuper{a,b}}IRIF, CNRS and Univ Paris-Diderot, France}
\email{\{buccia,kesner\}@irif.fr} 

\author[Simona Ronchi Della Rocca]{Simona Ronchi Della Rocca{\rsuper{c}}}	
\address{{\lsuper{c}}Dipartimento di Informatica, Universit\`a di Torino, Italy}	
\email{ronchi@di.unito.it}  

\keywords{Lambda-calculus, Type-Assignement Systems, Non-idempotent Intersection Types, Inhabitation Problem}
\subjclass{F.4.1 Mathematical Logic: Lambda calculus and related systems, Proof theory. F.3.1 Specifying and Verifying and Reasoning about Programs: Logics of programs}

\dedicatory{In honour of Furio Honsell, in occasion of his 60th birthday}
\begin{document}

\maketitle
\begin{abstract}
The inhabitation problem for intersection types in $\l$-calculus is
known to be undecidable. We study the problem in the case of
non-idempotent intersection, considering several type assignment
systems, which characterize the solvable or the strongly normalizing
$\l$-terms. We prove the decidability of the inhabitation problem for
all the systems considered, by providing sound and complete inhabitation
algorithms for them.
\end{abstract}


\begin{center}
\section{Introduction}
\end{center}
\medskip

Given a type assignment system associating types
to terms of a given programming language, two problems naturally
arise, namely the {\it typability} and the {\it inhabitation problem}
(known in the literature also as {\it emptiness problem}). In the
former, given a program, one wants to know if it is possible to assign
a typing to it, in the latter, in some specular way, given a typing,
one aims for a program to which the typing can be assigned. Since
types are program specifications, the decidability of the first
problem supplies tools for proving program correctness, whereas the
decidability of the second one provides tools for program
synthesis~\cite{Rehof15}. Considering the $\l$-calculus as a general paradigm for
functional programming languages, a virtuous example is the simple
type assignment system, for which both problems are decidable, and
which is the basis of typed functional languages, like ML and
Haskell. In this paper we study the inhabitation problem for an
extended type assignment system, based on (non-idempotent)
intersection types.

Intersection types have been introduced in order to increase the
typability power of simple type assignment systems, but quite early
they turned out to be a very powerful tool for
characterizing semantic properties of $\l$-calculus, like  solvability
and strong
normalization, and for describing models of
$\l$-calculus in various settings.  Intersection types have been
presented in the literature in many variants. Historically, one of the
first versions is the one characterizing solvable
terms~\cite{CoppoDezani:NDJoFL-80,krivine93book}, that we call system
$\cal C$, shown in Figure~\ref{fig:C}. In such a system, $\alpha$
denotes any basic type and the universal type is denoted by the
constant~$\omega$. Intersection enjoys associativity, commutativity,
and idempotency ($\A \inter \A =
\A$). Typing environments $\Gamma$ are functions from variables
  to types, we represent them as lists of pairs of the form $\x:\A$,
  where the comma symbol is used as set constructor.  System $\cal C$
assigns types different from $\omega$ to all and only those terms having head-normal forms,
which are the syntactical counterpart of the semantic notion of {\it
  solvability}~\cite{barendregt84nh}. Since it is undecidable to know
if a given term is solvable~\cite{barendregt84nh}, then
typability in system ${\cal C}$ (with types different from~$\omega$)
 is undecidable too. Moreover, the
inhabitation problem for system ${\cal C}$ has been proved to be
undecidable by Urzyczyn~\cite{Urzyczyn99}. Remark that system ${\cal C}$
is not syntax directed, so it is difficult to reason about it.  Van
Bakel~\cite{Bakel92} simplified system ${\cal C}$ by using strict types,
where intersection is not allowed on the right-hand side of the arrow; his
system ${\cal C}_B$ is presented in Figure~\ref{fig:S}, where 
we represent intersection 
through a set constructor, and
so the universal type $\omega$ as the empty set.
System ${\cal C}_B$
is syntax directed, \ie\ there are just three typing rules,
corresponding to the three different constructors of the
$\l$-calculus.
The systems
${\cal C}$ and  ${\cal C}_B$  have the same typability 
power, 
neglecting the universal type
$\omega$, in the sense that a term is typable in ${\cal C}$
by a type different form $\omega$ if and only if it is typable in ${\cal C}_B$.
In particular,  Urzyczyn's proof of undecidability of the inhabitation
problem for system ${\cal C}$ can be easily adapted to system ${\cal
C}_B$, proving that the latter is undecidable
too~\cite{UrzyczynPC}. 

\begin{figure}[!ht]
\[ \begin{array}{ccc}
\infer{\mbox{ }}{\Gamma \der \s : \omega} (\mathtt{\omega})&
\infer{\x:\A \in \Gamma}{ \Gamma\vdash \x : \A} (\axvarw)
& \infer
      {\Gamma, \x:\A \der \s : \D}
      {\Gamma \der \lambda \x.\s : \A \arrow \D}  
(\arri)\\ &&\\
\infer{
\Gamma\der \s : \A \arrow \D \quad
\Gamma \der \uu : \A}{\Gamma \der \s\uu : \D } (\arre)&
\infer{\Gamma \der \s : \A
\quad \Gamma \der \s : \D}{\Gamma \der \s : \A \inter \D}
 (\inti)& 
\infer {\Gamma\der \s :{{\A}_1} \inter {\A}_2} 
       {\Gamma \der \s : {{\A}_i} \ (i=1,2)} (\inte)\\ \\
\end{array}\] 
\begin{tabular}{l@{\hspace{.9cm}}l}
Types: & Typing environments:\\
$\A::= \alpha \mid \omega \mid \A\arrow \A \mid \A\inter\A$ & 
$\Gamma ::= \es \mid \Gamma,  \x:\A\ (\x\not\in\dom\Gamma)$
\end{tabular}
\caption{System ${\cal C}$}
\label{fig:C}
\end{figure}

\begin{figure}[!ht]
\[\begin{array}{c}
\infer{\x:\A \in \Gamma \sep \sigma\in\A}{ \Gamma \der \x : \sigma} (\axvarw) \quad \quad 
\infer{\Gamma, \x:\A\der \s : \tau}{\Gamma\der \lambda \x.\s : \A \arrow \tau}  (\arri) \\ \\
\infer{
\Gamma\der \s :\!\set{\sigma_i}_{i\in I} \arrow \tau \quad
(\Gamma \der \uu : \sigma_i)_{i\in I}} {\Gamma \der \s\uu : \tau  } (\arre) \\
\end{array}\]

\[\begin{array}{l@{\hspace{2cm}}l }
\mbox{Types:} & \mbox{Typing environments:} \\
\sigma,{\tau}::= \alpha   \mid \A\arrow \sigma\ (\mbox{strict types})
& \Gamma::=\es \mid \Gamma , \x:\A\ (\x\not\in\dom\Gamma)\\
\A::= \es \mid \iset{\sigma}\mid \A\cup\A\ (  \mbox{set types}) \\
\end{array}\]
\caption{System ${\cal C}_B$}
\label{fig:S}
\end{figure}

In this paper we study the inhabitation problem for {\it
  non-idempotent intersection types}, \ie\ considering intersection
modulo commutativity and associativity but not idempotence.
  It is possible to design various type
  assignment systems using non idempotent intersection;  we start
  from a system, that we call $\Mu$, which has been introduced in
  \cite{Gardner94} and further used by De Carvalho
  \cite{DeCarvalhoThesis}, for the purpose of studying the complexity
  of reduction. This system is particularly interesting since it
  induces a denotational model of $\lambda$-calculus, in the
  relational semantics setting \cite{PPR17}.  Moreover, it enjoys
  relevance and has a quantitative flavor.  We prove that the
  inhabitation problem is decidable by exploiting the fact that types
  keep track faithfully of the different uses of variables in terms,
  thanks to the relevance of the system.
 System $\Mu$ characterizes terms having head normal form.
 An important result for defining the algorithm solving 
the inhabitation problem for system $\Mu$  is an  approximation theorem, proved in section~\ref{s:approximate-lambda},
saying that a term can be assigned all and only the types  that can be assigned to its approximants.
We solve the inhabitation problem for system $\Mu$ in a constructive way, by designing a sound and complete algorithm, that
given a typing environment $\Gamma$ and a type $\sigma$, builds a set
of approximate normal forms from which all and only the head-normal
forms $\s$ such that $\Gam \der \s:\sigma$ can be generated. 
Then we
extend the system with a weakening rule, yielding system $\Mu_w$, and we prove that
inhabitation remains decidable for such an extension.

In the second part of this paper
we  present some  non-idempotent intersection types systems characterizing strong
normalization, and we show that the inhabitation problem is decidable for each of them. 
We take into account different systems present in literature, also when they do not enjoy good properties like subject reduction,  
with the aim of performing a complete analysis.

\paragraph{\bf Interest of the problem} As said before, inhabitation
has been used for program synthesis (see for example~\cite{Rehof15}),
and it is a technical problem interesting by itself.  The results of
this paper show in particular that the two classical problems about
type assignment systems, namely typability and inhabitation, are
unrelated problems, giving a first example of a system for which the
first one is undecidable while the second one is decidable. This fact
suggests interesting questions about the relation between program
correctness and program synthesis.

\paragraph{\bf Related work} Various restrictions of the 
standard
intersection type system have been shown to have decidable
inhabitation problems~\cite{KurataT95,Urzyczyn10}. The approach is
substantially different from the one used in this work, since in all
cases intersection is idempotent, and the decidability is obtained by
restricting the use of rules $(\inti)$ and $(\inte)$, so that the
corresponding type assignment system does not characterize interesting
classes of terms, anymore.  Recently a new notion of intersection type
system with bounded dimension has been
introduced~\cite{DBLP:conf/popl/DudenhefnerR17}, it is based on
decorations of terms called elaborations, that remember some
information about their typing derivation, and allow a stratification
of typed terms. It has been proved that inhabitation (where
inhabitants are decorated terms) is decidable, and its complexity is
EXPSPACE complete. Previously it had been proved in~\cite{Kus07} that  the inhabitation of rank two intersection types is EXPTIME hard.

A preliminary version of our inhabitation algorithm for system $\Mu$ has
been presented in~\cite{DBLP:conf/ifipTCS/BucciarelliKR14}. The present paper simplifies this
first algorithm (see discussion in Section~\ref{s:inhabitation}) and extends inhabitation to other systems, as explained before.

\paragraph{\bf Non-idempotent intersection types} In the last years, growing interest has been devoted to 
non-idempotent
intersection types, since they allow to reason about quantitative
properties of terms, both from a syntactical and a semantic point of
view.  In fact, system $\Mu$ is not new: it is the system of
Gardner~\cite{Gardner94} and de Carvalho~\cite{DeCarvalhoThesis}, and
it is an instance of the class of the essential $\lambda$-models
defined in~\cite{PPR17}, which supplies a logical description of the
strongly linear relational $\lambda$-models.  Some other type
assignment systems with non-idempotent intersection have been studied
in the literature, for various purposes: to compute a bound for the
normalization time of terms~\cite{DebeneRonchiITRS12}), to supply new
characterizations of strong normalization~\cite{BernadetL13,KV14}, to
study type inference~\cite{kfouryWells04,MairsonNeergaard04}, to study
linearity~\cite{kfoury00}, to characterize solvability in the resource
$\lambda$-calculus~\cite{pagani10fossacs,PaganiRonchi:2010FI},
to characterize the set of hereditary head-normalizing infinite $\l$-terms~\cite{Vial17}.
Moreover intersection without idempotency, commutativity nor
associativity, has been used to study the game semantics of a typed
$\lambda$-calculus~\cite{DiGia08}.  Non-idempotent types
are also useful to prove observational equivalence of
programming languages~\cite{Kesner16,BalabonskiBBK17}. 
A unified model-theoretical approach
covering both the relevant and non-relevant cases, and unveiling the
relations between them, is presented in~\cite{Ehrhard12}.  
Non-idempotent (intersection and union) types have also been
proposed~\cite{KV17} to characterize different operational properties of 
the $\lambda\mu$-calculus, a computational interpretation of 
classical logic in  natural deduction style. A survey
on non-idempotent intersection type assignement systems for
the $\l$-calculus and the proof techniques for them can be found
in~\cite{BKV17}.

\paragraph{\bf Organization of the paper} Section~\ref{sec:prelim}  contains preliminary notions about the {$\l$-calculus}; 
Section~\ref{sec:hn} presents the system $\Mu$ and its inhabitation algorithm; Section~\ref{section:sn} presents the systems
$\Mu_{e,w}$ and $\Su_w$, both characterizing strong normalisation, and
their respective inhabitation algorithms. Finally,
Section~\ref{sec:concl} proposes some conclusions.

\section{Preliminaries}\label{sec:prelim}
\paragraph{\bf The $\l$-calculus} Terms and contexts of the $\l$-calculus are generated by the following grammars, respectively:
\[  \s,\uu,\vv ::= \x \mid  \l\x.\s \mid \s\uu \quad \quad \quad \quad 
\ccontext  ::=  \square \mid \l \x. \ccontext  \mid  \ccontext\s \mid \s\ccontext 
 \] 
where $\x$ ranges over a countable set of variables. 
As usual, the application symbol associates
  to the left;  to avoid ambiguities in the notation
  we may use parentheses  as \eg\ in $\x (\x \y)$.  We use the
notation $\iid$ for the identity function $\l \x. \x$ and $\dup$ for
$\l \x. \x \x$, which duplicates its argument.  We
write $\fv{\s}$ to denote the set of free variable
of $\s$ and $=$ for the syntactical equality on terms, modulo renaming
of bound variables.  The notation $\l \x\y.\s$ is used as an
abbreviation for $\l\x.\l\y.\s$. 
We assume an
  hygiene condition on variables, \ie\ free and bound variables have
  different names, as well as  variables bound by different
  binders.
Given a context $\ccontext$ and a term $\s$, $\ccontext[\s]$ denotes
the term obtained by replacing the unique occurrence of $\square$ in
$\ccontext$ by $\s$, thus potentially allowing the capture of free variables of $\s$.
 A context $\ccontext$ is closing for $\s$ if
  $\ccontext[\s]$ is closed, \ie\  if $\fv{\ccontext[\s]}=\es$.

The $\beta$-reduction, denoted by $\redb$, is the contextual closure of the rule:  
\[(\l \x.\s)\uu\arrow \s\isubs{\uu}{\x}\] 
where $\s \isubs{\uu}{\x}$ denotes the capture-free replacement of $\x$ by $\uu$ in $\s$.
A term of the form $(\l \x.\s)\uu$ is called a $\beta$-redex. We use $\redbs$ to denote the reflexive and transitive closure of $\redb$, and $=_\beta$ the transitive, reflexive and symmetric closure of $\redb$.

\paragraph{\bf Normal forms} A term $\s$ {\it is in $\beta$-normal form}, 
or just in normal form (\nf), when it does not contain any
$\beta$-redex, it {\it has normal form} if it can be reduced to a term in
normal form, it is {\it strongly normalizing}, written $\s \in SN$,  if every $\beta$-reduction
sequence starting from it eventually stops. Terms in normal form (${\mathcal F}$) are
generated by the following grammar:
\[{\cal F} ::=  \l \x. {\cal F} \mid {\cal H} \quad \quad\quad\quad\quad
  {\cal H}   ::=  \x   \mid {\cal H} {\cal F} \] 

\paragraph{\bf Head-normal forms}
The notion of head-normal form (\hnf) is the syntactical counter part
of the well-known notion of solvability for the $\lambda$-calculus,
being a term $\s$ solvable iff there is a closing context
$\ccontext $ for $\s$ of the shape $(\l \x_1...\x_n. \square) \s_1...\s_m$ (called a head-context) such
that $\ccontext[\s] =_\beta \iid$.  A
$\l$-term {\it is in \hnf} if it is generated by the following grammar
${\cal J}$, it {\it has \hnf} if it $\beta$-reduces to a term which is in \hnf.
\[  {\cal J}  ::=  \l \x. {\cal J}  \mid  {\cal K} \quad \quad\quad\quad\quad
    {\cal K}   ::=  \x   \mid {\cal K} \s \] 

\noindent For example, $\s= (\l \y. \y)\x \dup$ is not in \hnf\ but 
$\beta$-reduces to $\x \dup$ which is in \hnf, thus $\s$ has a \hnf. 
The term $\iid$ is solvable while $\dup \dup$ is not. 

\paragraph{\bf Approximate normal forms}
Approximate normal forms~\cite{barendregt84nh} are normal forms in an extended 
calculus.
Let $\Lambda\Omega$ be the $\lambda$-calculus enriched with  a constant $\Omega$, and let $\redo$ be the contextual closure of the $\beta$-reduction plus the following two reduction rules:
\[\Omega \s \arrow \Omega \quad \quad\quad\quad\quad \l\x.\Omega \arrow \Omega  \]
Normal forms of $\Lambda\Omega$  with respect to $\redo$ are defined through the following grammar: 
\[ {\ap},{\bp},{\cp} \mbox{ } ::= \mbox{ }\Omega \mid   {\cal N}  \qquad
   {\cal N}  \mbox{ } ::= \mbox{ }\l \x. {\cal N} \mid {\cal L}
                  \qquad {\cal L} \mbox{ } ::= \mbox{ }\x \mid  {\cal L} \ap\\
 \]

\noindent Elements generated by  the three grammar rules above are called 
  approximate normal forms, ${\cal N}$ approximate normal forms and ${\cal L}$ approximate normal forms, respectively.
E.g. both $\l \x \y. \x \Omega (\l z. \y\z\Omega)$ and $\l \x \y. \x \Omega \Omega$ are approximate normal forms, while 
$\l \x.\Omega$ and $\l \x. \x \y(\Omega \y)$ are not, since they reduce respectively to $\Omega$ and $\l \x. \x \y\Omega$. 


\paragraph{\bf Approximants of a term}
 Approximate normal forms 
can be ordered by  the smallest contextual  
order $\leq$ such that $\Omega \leq \ap$, for all $\ap$.
By abuse of notation, we write $\ap \leq \s$ to compare an approximate normal form 
$\ap$
with a term $\s$ when  $\s$ is obtained from $\ap$ by
replacing all the occurrences of $\Omega$ by arbitrary terms. Let $\bigvee$ denote the least upper bound w.r.t. $\leq$.
We use the predicate $\uparrow_{\iI}\ap_i$ to state that $\bigvee\set{\ap_i}_{\iI}$
does exist.

The {\it set of approximants of a term} $\s$ is given by:
\[\Ap(\s)=\set{\ap \ |\ \exists \uu\ \s\arrow_\beta^*\uu \mbox{ and }\ap\leq \uu}\]
It is easy to check that, for every $\s$ and
$\ap_1,\ldots \ap_n \in \Ap(\s)$,
$\uparrow_{i\in\set{1,\ldots,n}}\ap_i$. 
Thus for example, $ \Ap(\l \x. \dup(\iid\iid)) = \set{\Omega, \l
  \x. \Omega, \l \x\y. \Omega, \l \x\y.\y}$ and $\bigvee \Ap(\l \x. \dup(\iid\iid))=\l \x\y.\y$.

 In many well-behaved
models, it is possible to relate the interpretation of
a term to the interpretation of its approximants, the first example being in \cite{wadsworth76siamjc}.  We are going to show 
such a property, known as the
{\it approximation theorem}, for the type assignment system
$\Mu$ presented below.

\section{Systems characterizing head normalization}\label{sec:hn}
In this section we consider the inhabitation problem with respect to
an intersection type system for the $\lambda$-calculus
that characterizes 
head-normal forms.  The essential feature of this type system that makes its inhabitation problem decidable is the non-idempotency of the intersection.
The typing rules are relevant (\ie\ the weakening is not allowed), 
and we represent
non-idempotent intersections as multisets of types. Moreover,
in order to work with a syntax-directed system, we 
restrict the set of types to the {\it strict ones}~\cite{Bakel92}. 

\begin{defi}\label{def::lambdasys}\leavevmode
\begin{enumerate}
\item The set of {\it types}  is defined by the following grammar:
\[ \begin{array}{llllll}
\sigma, \tau, \rho & ::= & \alpha \mid \A \arrow \tau & (\mbox{types})  \\  
\A, \D  & ::= & \multiset{\sigma_{i}}_{\iI}  &  (\mbox{multiset types}) 
\end{array} \] 
where $\alpha$ ranges over a countable set of base types.
Multiset types are associated to a finite, 
possibly empty, set $I$ of indices;  the {\it empty multiset} corresponds to the case $I=\emptyset$
and  is simply denoted by $\emm$. To avoid an excessive number of parentheses,
we  use for example $\A \arrow \D \arrow \sigma$ instead of $\A \arrow (\D \arrow \sigma)$.
\ignore{ \item The  non-deterministic
{\it choice} operator on multiset types is defined as follows:
\[\A^*   := \left \{ \begin{array}{lllll}
                      \A & \mbox{ if } \A \neq \emul \\
                      \multiset{\tau}  &  \mbox{ if } \A = \emul,  \mbox{ where } \tau \mbox{ is an arbitrary type} \\
\end{array}  \right . \]
This operator will play a special role in Section~\ref{section:sn}. }

\item {\it Typing environments}, which are also simply called environments, written
  $\Gamma, \Delta$, are functions from variables to multiset types,
  assigning the empty multiset  to almost all the
  variables. We use the symbol $\es$ to denote the empty typing environment. The domain of $\Gamma$, written $\dom{\Gam}$, is the set
  of variables whose image is different from $\emul$. Given
  environments $\Gamma$ and $\Delta$, $\Gamma + \Delta$ is the
  environment mapping $\x$ to $\Gam (\x)\uplus\Delta(\x)$, where
  $\uplus$ denotes multiset union; $+_{\iI} \Delta_i$ denotes  its obvious
  extension to a non-binary case, where the resulting environment 
  has empty domain in the case  $I=\es$.
We write $\Gamma\sm \x$ for 
the environment assigning $\emul$ to $\x$, and acting as $\Gamma$
otherwise; ${\x}_1:{\A}_1,\ldots,{\x}_n:{\A}_n$ is the environment
assigning ${\A}_i$ to ${\x}_i$, for $1\leq i\leq n$, and $\emul$ to
any other variable. 

\item A {\it typing judgement} is a triple either of the form $\Gam \vdash \s:\sig$ or $\Gam \vdash \s:\A$.
The type system $\Mu$ is given in Figure~\ref{fig:system-lambda}. 
\end{enumerate}
\end{defi}

\begin{figure}[!ht]
\begin{center}
$\begin{array}{c}
\infer{\mbox{}}{\x:\multiset{\rho} \der \x:\rho}\ (\axvar)
\quad \quad 
\infer
{\Gam\der \s:\tau }
{\Gam \sm  \x \der \l \x. \s: \Gamma(\x)\arrow \tau}\ (\arri)\\ \\
\infer{(\Delta_{i} \der \s:\sig_i)_{\iI}} {+_{\iI} \Delta_{i} \der \s:\multiset{\sig_{i}}_{\iI}}(\many)   \quad \quad 
\infer
{\Gam\der \s:	\A\arrow\tau \sep 
\Delta\der \uu:\A }
{\Gam +\Delta\der \s\uu: \tau}\ (\arre)
\end{array}$
\caption{The type assignment system $\Mu$ for the $\l$-calculus}
\end{center}
\label{fig:system-lambda}
\end{figure}

\noindent Rules $(\axvar)$ and $(\arri)$ are self explanatory.  Rule $(\many)$
can be considered as an auxiliary rule, \ie\ it has no logical meaning
but just collects together different typings for the same term; remark
that it cannot be iterated.  Rule $(\arre)$ has two premises, the one
for $\s$ (resp. $\uu$) is called the major (resp. minor) premise. In
the case $\A = \emul$, this rule allows to type a term without giving
types to all its subterms, and in particular it allows to type
an application whose argument is unsolvable.  
For example, the judgement $\x:\multiset{\emul\arrow\alpha}\der\x(\dup
\dup):\alpha$ turns to be derivable by taking $I =\emptyset$ in  rule $(\many)$ :

\[ \begin{prooftree}
\begin{prooftree}
\justifies{\x:\multiset{\emul\arrow\alpha}\der\x: \emul\arrow\alpha}
\using{(\axvar)}
\end{prooftree} \sep
\begin{prooftree}
\justifies{\emptyset \der \dup\dup :\emm }
\using{(\many)}
\end{prooftree}
\justifies{\x:\multiset{\emul\arrow\alpha}\der\x(\dup\dup):\alpha}
\using{(\arre)}
\end{prooftree} \] 
This
feature is shared by all the intersection type systems characterizing
solvability.

\begin{nota}
We write $\Pi \dem \Gam
\der_{\Mu} \s:\sigma$, or simply $\Pi \dem \Gam \der \s:\sigma$ when
$\Mu$ is clear from the context, to denote a type
  derivation $\Pi$ in system $\Mu$ with
    conclusion $\Gam \der \s:\sigma$.
We call $\s$ the {\it subject} of $\Pi$.  
A term $\s$ is said to be $\Mu$-typable if there exists a derivation $\Pi \dem \Gam
\der_{\Mu} \s:\sigma$. By abuse of notation, we omit  sometimes
the name of the derivation by writing simply  $\Gam \der
\s:\sigma$ to denote the existence of some derivation with
conclusion $\Gam \der \s:\sigma$.  We  extend these
notations to judgements of the shape $\Gam \der \s: \A$, when we want
to reason by induction on the structure of a proof.  
\end{nota}

One important tool is  going to be used in the proofs: 
\begin{defi} \label{meas-sub} \mbox{} 
The {\it measure} of a derivation $\Pi$, written $\meas(\Pi)$, 
is the  number of rule applications  in $\Pi$, except  rule $(\many)$.
\end{defi}

Note that the definition of the measure of a type derivation reflects
the fact that $(\many)$ is an auxiliary rule.  The notion of measure of a derivation provides an
original, combinatorial proof of the fact that typed terms do have
{\hnf}.  In fact the fundamental {\it subject reduction} property
holds as follows: if $\Pi\dem\Gamma \der\s:\sigma$ and
$\s\arrow_\beta\uu$, then $\Pi'\dem\Gamma \der\uu:\sigma$, with the
peculiarity that the measure of $\Pi'$ is strictly smaller than that
of $\Pi$ whenever the reduction $\s\arrow_\beta\uu$ takes place in a
subterm  of $\s$ which is {\it typed in $\Pi$}. A formal definition
of typed positions  follows.

\begin{defi}
  \label{def:occ}\mbox{}
\begin{itemize}

\item The set $\occ t$ of positions of  $\s$ is  the set of contexts
$\ccontext$ such that there exists a term $\uu$ verifying 
 $\ccontext[\uu]=\s$, $\uu$ being the {\it  subterm of $\s$  at position $\ccontext$}.

\item Given $\Pi\dem\Gamma\der \s:\sigma$, the {\it set $\tocc\Pi\subseteq\occ \s$ of
  typed positions of  $\s$  in  $\Pi$} is defined by induction on 
the structure of $\Pi$ as follows:
\begin{itemize}
\item $\tocc\Pi=\set{\square}$ if $\Pi$ is an instance of the axiom.
\item $\tocc\Pi=\set{\square}\cup \set{\l\x.\ccontext\ | \ \ccontext\in \tocc{\Pi'}}$ if the 
last rule of $\Pi$ is $(\arri)$, its subject is $\l \x. \uu$  and its premise is $\Pi'$.
\item $\tocc\Pi=\set{\square}\cup \set{\ccontext\vv\ | \ \ccontext\in \tocc{\Pi'}}
\cup \set{\uu\ccontext\ |\ \ccontext\in\tocc{\Delta}}$  if the 
last rule of $\Pi$ is $(\arre)$, its subject is $\uu \vv$, $\Pi'$ and $\Delta$ are  the major and minor premises of $\Pi$ respectively.
\item $\tocc\Pi= \bigcup_{\iI}\set{\ccontext\ |\ \ccontext\in\tocc{\Pi'_i}}$ if the 
  last rule of $\Pi$ is $(\many)$, and  $(\Pi'_i)_{\iI} $ are its  premises.
A particular case is when $I = \es$, in which case
    $\tocc \Pi = \es$. This coincides with the fact that terms typed by an empty
    $(\many)$ rule are semantically untyped.
\end{itemize}

We say that {\it the subterm of $\s$ at position $\ccontext$
  is typed in $\Pi$} if $\ccontext$ is a typed 
position of $\s$ in $\Pi$.
\item  Given $\Pi\dem\Gamma\der \s:\sigma$ (resp. $\Pi\dem\Gamma\der \s:\A$), we say that $\s$ is {\it in $\Pi$-normal form},
written $\Pi\demnf\Gamma\der \s:\sigma$ (resp. $\Pi\demnf\Gamma\der \s:\A$),
if for all $\ccontext\in\tocc\Pi$, $\s= \ccontext[\uu]$ implies
$\uu$ is not a redex.
\end{itemize}
\end{defi}
 
\begin{exa}\label{ex:meas}
Consider the following derivation $\Pi$, 
where $\sig:= \multiset{\alpha_0, \alpha_1} \arrow \emul \arrow \tau$:

\[ \begin{prooftree}
   \begin{prooftree}
   \begin{prooftree}
   \justifies{ \x: \multiset{\sig} \der \x: \sig}
   \using{(\axvar)}
   \end{prooftree} 
   \begin{prooftree}
   \begin{prooftree} \justifies{\y: \multiset{\alpha_0} \der \y:\alpha_0 }\using{(\axvar)}\end{prooftree} 
   \begin{prooftree} \justifies{\y: \multiset{\alpha_1} \der \y:\alpha_1 }\using{(\axvar)}\end{prooftree} 
   \justifies{\y: \multiset{\alpha_0, \alpha_1}  \der \y: \multiset{\alpha_0, \alpha_1}  }
   \using{(\many)}
   \end{prooftree} 
   \justifies{\x: \multiset{\sig}, \y: \multiset{\alpha_0, \alpha_1} \der \x \y: \emul \arrow \tau} 
   \using{(\arre)}
   \end{prooftree} 
     \begin{prooftree}
      \justifies{\es \der \iid\dup: \emul}\using{(\many)}
      \end{prooftree}
   \justifies{\x: \multiset{\sig}, \y: \multiset{\alpha_0, \alpha_1} \der \x \y (\iid\dup) : \tau}
   \using{(\arre)}
   \end{prooftree} \] 

\noindent Then, 
\begin{itemize}
\item$\meas(\Pi) = 5$.
\item $\tocc{\Pi} = \set{\Box, \Box\y (\iid\dup), \x \Box (\iid\dup), \Box (\iid\dup)} $.
\item $\x \y (\iid \dup)$ is in $\Pi$-normal form (since its only redex $\iid \dup$ is untyped). 
\end{itemize}
\end{exa}

\begin{thm}\label{thr::char}\leavevmode
\begin{enumerate}
\item \label{th::red1}(Subject reduction and expansion) $\Gamma \der \s:\sigma$ and $\s =_\beta \uu$ imply $\Gamma \der \uu:\sigma$.
\item \label{th::red2} (Characterization) $\s$ is $\Mu$-typable if and only if $\s$ has  \hnf.
\end{enumerate}
\end{thm}
\proof 
See~\cite{DeCarvalhoThesis, BKV17}. In particular, the proof of
subject reduction is based on the weighted subject reduction property that $\Pi \dem \Gamma \der
\s:\sigma$ and $\s \redb \uu$ imply $\Pi'\dem\Gamma \der \uu : \sigma$,
where $\meas(\Pi') \leq \meas(\Pi)$. Moreover, if the reduced redex 
is typed in $\Pi$, then $\meas(\Pi') <\meas(\Pi)$.
\qed

As a matter of fact, the two properties stated in the theorem above
may be proved using a semantic shortcut: in~\cite{PPR17} the class of
{\it essential} type systems is introduced, and it is shown that such
systems supply a logical description of the linear relational models
of the $\l$-calculus, in the sense of~\cite{bucciarelli07csl}. Since the
type system $\Mu$ is an instance of this class, both statements of
Theorem~\ref{thr::char} are particular cases of the results proved in~\cite{PPR17}.

\subsection{The key role of approximants}
\label{s:approximate-lambda}

System $\Mu$ assigns  types to  terms without giving types to all their
subterms (\cf\ the rule $(\arre)$ in case $\A=\emul$). So in order to
reconstruct all the possible subjects of  derivations we
need a notation for these untyped subterms, which is supplied by the
notion of approximate normal forms introduced
  in Section~\ref{sec:prelim}.
As a consequence,
  some notions previously defined for terms, naturally apply for
  approximants too. Thus for example, given $\Pi\dem\Gamma\der
  \ap:\sigma$ or $\Pi\dem\Gamma\der \ap:\A$, $\ap$
  is said to be in $\Pi$-normal form according to Definition~\ref{def:occ}.
But remark that {\it every} typing derivation of an approximant is
  necessarily in normal form.  More precisely, if $\Pi\dem\Gamma\der
  \ap:\sigma$ or $\Pi\dem\Gamma\der \ap:\A$, then $\ap$ is in $\Pi$
  normal form.  Quite surprisingly, $\Pi\dem\Gamma\der \ap:\sigma$,
  $\ap\leq \s$ and $\Pi'\dem\Gamma\der \s:\sigma$ do not imply that
  $\s$ is in $\Pi'$ normal form, as the following example shows.

\begin{exa}
Let $\ap = \x \Omega (\x \y \y) \leq  \x (\iid \x) (\x \y \y) = \s$. 
Let $\Gam = \x: \multiset{\sig_1, \sig_2}, \y: \multiset{\alpha,\alpha}$, 
where $\sig_1 = \emm \arrow \multiset{\alpha} \arrow \alpha$
and $\sig_2 = \multiset{\alpha} \arrow \multiset{\alpha} \arrow \alpha$.
There are type derivations $\Pi \dem \Gam \der \ap: \alpha$ (given on the top) and
 $\Pi' \dem \Gam \der \s: \alpha$ (given in the bottom) such that
$\ap$ is in $\Pi$-normal form but $\s$ is not in
$\Pi'$-normal form.

\[  \begin{prooftree}
    \begin{prooftree} 
   \begin{prooftree} 
   \justifies{\x: \multiset{\sig_1} \der \x: \sig_1 }
   \end{prooftree} \sep
   \begin{prooftree} 
   \justifies{\emptyset \der \Omega:\emm }
   \end{prooftree} 
   \justifies{\x: \multiset{\sig_1} \der \x \Omega: \multiset{\alpha} \arrow \alpha }
   \end{prooftree} \sep
   \begin{prooftree} 
   \begin{prooftree} 
   \begin{prooftree} 
   \begin{prooftree} 
   \justifies{\x: \multiset{\sig_2} \der \x:\sig_2}
   \end{prooftree} \sep
   \Pi_\y 
   \justifies{\x: \multiset{\sig_2}, \y: \multiset{\alpha} \der \x \y:  \multiset{\alpha} \arrow \alpha}
   \end{prooftree} \sep 
   \Pi_\y
   \justifies{\x: \multiset{\sig_2}, \y: \multiset{\alpha, \alpha} \der \x \y \y: \alpha}
   \end{prooftree}   
   \justifies{\x: \multiset{\sig_2}, \y: \multiset{\alpha, \alpha} \der \x \y \y: \multiset{ \alpha}}
   \end{prooftree}   
   \justifies{\x: \multiset{\sig_1, \sig_2}, \y: \multiset{\alpha, \alpha} \der \x \Omega (\x \y \y): \alpha}
   \end{prooftree}    \]

\[ \begin{prooftree}
    \begin{prooftree} 
   \begin{prooftree} 
   \justifies{\x: \multiset{\sig_2} \der \x: \sig_2 }
   \end{prooftree} \sep
   \begin{prooftree} 
   \begin{prooftree}
    \begin{prooftree}
    \begin{prooftree}
   \justifies{\z: \multiset{\alpha}  \der \z:  \alpha  }
   \end{prooftree}
   \justifies{\emptyset \der \iid: \multiset{\alpha} \arrow \alpha   }
   \end{prooftree} \sep
   \Pi_\y
   \justifies{\y: \multiset{\alpha} \der \iid \y: \alpha }
   \end{prooftree} 
   \justifies{\y: \multiset{\alpha} \der \iid \y: \multiset{\alpha} }
   \end{prooftree} 
   \justifies{\x: \multiset{\sig_2},\y:\multiset{\alpha} \der \x (\iid \y): \multiset{\alpha} \arrow \alpha }
   \end{prooftree} \sep
   \begin{prooftree} 
   \begin{prooftree} 
   \begin{prooftree} 
   \begin{prooftree} 
   \justifies{\x: \multiset{\sig_1} \der \x:\sig_1}
   \end{prooftree} \sep
   \begin{prooftree} 
      \justifies{\emptyset  \der \y:  \emm  }
   \end{prooftree} 
   \justifies{\x: \multiset{\sig_1} \der \x \y:  \multiset{\alpha} \arrow \alpha}
   \end{prooftree} \sep 
   \Pi_\y 
   \justifies{\x: \multiset{\sig_1}, \y: \multiset{\alpha} \der \x \y \y: \alpha}
   \end{prooftree}   
   \justifies{\x: \multiset{\sig_1}, \y: \multiset{\alpha} \der \x \y \y: \multiset{ \alpha}}
   \end{prooftree}   
   \justifies{\x: \multiset{\sig_1, \sig_2}, \y: \multiset{\alpha, \alpha} \der \x (\iid \y) (\x \y \y): \alpha}
   \end{prooftree}    \] 

where  $\Pi_\y$ denotes the following subderivation
\[  \begin{prooftree} 
   \begin{prooftree}
   \justifies{ \y:  \multiset{\alpha}    \der \y:  \alpha   }
   \end{prooftree}
   \justifies{ \y:  \multiset{\alpha}    \der \y:  \multiset{\alpha}   }
   \end{prooftree}\] 
\end{exa}

\noindent Given
$\Pi\dem \Gam \der \s:\tau$, where $\s$ is in $\Pi$-normal form, we
denote by $\Ap(\Pi)$ the least approximate normal form $\bp$ of $\s$ such that
$\Pi\dem \Gam \der \bp:\tau$.   Formally:

\begin{defi}\label{def:approx}
Given $\Pi\demnf\Gam \der \s:\sig$, the approximate normal form 
$\Ap(\Pi)\in\Ap(\s)$, called the {\it approximant of $\Pi$},  is defined by induction on the structure of $\Pi$ as follows:
\begin{itemize}
\item If the last rule of $\Pi$ is $(\axvar)$, then $\Pi\dem \Gam \der \x:\sig$, and $\Ap(\Pi) := \x$.
\item If the last rule of $\Pi$ is $(\arri)$, then $\sigma= \A \arrow \rho$ and $\Pi\dem\Gam \der \l \x. \s: \A \arrow \rho$ follows  from 
     $\Pi' \dem\Gam,\x:\A \der \s:\rho$, then  $\Ap(\Pi) := \lambda \x. \Ap(\Pi')$,  $\s$ being in  $\Pi'$-normal form.
\item If the last rule of $\Pi$ is $(\arre)$, $\Pi \dem \Gam + \Delta
  \der \vv\uu: \sig$ with premises $\Pi'\dem\Gam\der \vv: \A \to \sig$
  and $\Pi''\dem\Delta\der\uu:\A$, so by induction $\Ap(\Pi')
    \in \Ap(\vv)$ and $\\Ap(\Pi'') \in \Ap(\uu)$. Moreover the
  hypothesis that $\vv\uu$ is in $\Pi$-normal form implies that $\vv$ is of the form 
$\x\vv_1\ldots\vv_n$, so that
  $\Ap(\Pi):=\Ap(\Pi')\Ap(\Pi'') \in \Ap(\vv\uu)$.
\item  If the last rule of $\Pi$ is $(\many)$,  then $\Pi\dem\Gam \der \uu: \A $ 
follows from $(\Pi_i \dem\Gam_i \der \uu: \sig_i)_{\iI}$, then $\Ap(\Pi)=\bu_{\iI} \Ap(\Pi_i)$. In fact $\uparrow_{\iI}\Ap(\Pi_i )$, since $\Ap(\Pi_i )\in \Ap(\uu)$, for all $\iI$.
\end{itemize}
\end{defi}

Remark that in the last case the approximate normal form corresponding
to the case $I=\es$ is $\Omega$. Coming back to the derivation $\Pi$
in Example~\ref{ex:meas}, we have $\Ap(\Pi) = \x \y \Omega$. More
generally, given $\Pi\demnf\Gam \der \s:\sig$, the approximant $\Ap(\Pi)$ can be
obtained from $\s$ by replacing all its maximal subterms untyped in
$\Pi$ by $\Omega$.

Simple inductions on  $\Pi$ allow to show the following properties:
\begin{prop}\label{Prop:monoton}\mbox{}
\begin{enumerate}
\item\label{monone} Let $\Pi\dem\Gamma\der \ap:\sigma$.
  If  $\ap\leq\bp$  (resp. $\ap\leq\s$)  then there exists $\Pi'$ such that
$\Pi'\dem\Gamma\der \bp:\sigma$  (resp. $\Pi'\demnf\Gamma\der \s:\sigma$) and $\Ap(\Pi')=\Ap(\Pi)$.
\item\label{montwo}
If $\Pi \dem \Gam \der \ap:\sigma$ or $\Pi \demnf \Gam \der \s:\sigma$, then    there exists
$\Pi' \dem \Gam \der \Ap(\Pi):\sigma$ and $\Ap(\Pi')=\Ap(\Pi)$.

\end{enumerate}
\end{prop}

Proposition~\ref{Prop:monoton}.\ref{montwo} ensures the completeness of approximants, in the sense that 
all the typings having subject $\s$ may also 
type  some approximant of 
$\s$. 

On the other hand, it is easy to see that if $\Pi \dem \Gam \der \s:\sigma$ and $\s$ is in $\Pi$-normal form, then $ \Ap(\Pi)\leq \s$. Summing up, we get a new proof of the approximation theorem
for the relational $\lambda$-model associated to system $\Mu$, in the sense of~\cite{PPR17}:

\begin{thm}\label{th:approx}
The relational $\lambda$-model induced by system $\Mu$ satisfies the
approximation theorem, \ie\ the interpretation of a term is the union of the 
interpretations of its approximants.
\end{thm}

We are ready to prove that the set of approximate normal forms solving any given instance of the inhabitation problem is finite. 
This property allows us to design a complete inhabitation algorithm, \ie\  one which supplies all solutions.  
First, we will prove that a derivation in normal form enjoys a sort of subformula property. 
Let us define  the {\it subtype} relation as the transitive closure
of the following one:
\begin{itemize}
\item $\sig$ is a subtype of $\sig$.
\item $\A$ is a subtype of $\A$.
\item $\A$ and $\tau$ are subtypes of $\A \arrow \tau$.
\item $\sig_i$ is a subtype of $\multiset{\sig_i}_{\iI}$ for all $\iI$.
\end{itemize} 

Finally, $\A$ is a subtype of $\Gam$ if $\x:\A \in \Gam$.

\begin{lem}
\label{lem:subformula}
Let $\ap$ be an approximate normal form. 
Let $\Pi \dem \Gam \der
\ap:\sigma$. For every subderivation of $\Pi$ of the shape
$\Pi'\dem \Gam' \der \bp: \tau$, 
$\tau$ is either  a subtype of
$\sigma$ or a subtype of  $\Gam$. Moreover, if $\ap$ is 
an ${\cal L}$ approximate normal form, then $\sig$  is a subtype of $\Gam$.
\end{lem}

\proof The proof is by induction on $\Pi$. We do not consider the case of 
rule $(\many)$ since the derivation ends in a type $\sigma$.

If $\Pi$ ends with $(\axvar)$, 
then $\ap = \x$ and  $\Pi \dem \x:\multiset{\sig} \der \x:\sig$, and 
the property is obviously true, being $\sigma$ a subtype of itself.
Moreover, $\x$ is an ${\cal L}$ approximate normal form,
and $\sig$ is a subtype of $\multiset{\sig}$, thus a subtype 
of $\x:\multiset{\sig}$.

If $\Pi$ ends with $(\arri)$, 
then $\ap= \l \x.\ap'$,  $\sig= \A \arrow \tau$, and $\Pi$ has the following shape: 
\[\infer
{\Gam, \x:\A\der \ap':\tau }
{\Gam  \der \l \x. \ap': \A\arrow \tau}\ (\arri)\\ \\	\]
By the \ih\  
the property holds  for 
the derivation with subject $\ap'$.
Since both  $\A$ and $\tau$ are subtypes of $\A\arrow \tau$, we conclude by 
transitivity of the subtype relation. The moreover statement does not apply
to abstractions.

If $\Pi$ ends with $(\arre)$, 
then $\ap=\ap_{0}\ap_{1}$, where $\ap$ and  $\ap_0$ are ${\cal L}$ approximate
normal forms, and $\ap_1$ is an approximate normal form. Then $\Pi$ has the following shape:  
\[
\begin{prooftree}
\Gam\der \ap_{0}:	\A\arrow\tau 
\begin{prooftree}
(\Delta_i \der \ap_1:\rho_i)_{\iI}
\justifies{\Delta\der \ap_{1}:\A} \using{(\many)}
\end{prooftree}
\justifies{\Gam +\Delta\der\ap_{0}\ap_{1}: \tau}
\using{(\arre)}
\end{prooftree}
\]
where $\Delta = +_{\iI} \Delta_i$ and $\A = \multiset{\rho_i}_{\iI}$. 
We have that $\rho_i$ is a subtype of $\A$, which in turn  is a subtype of $\A \arrow \tau$,
and by the \ih\ 
$\A \arrow \tau$ is a subtype of $\Gam$, thus a subtype of $\Gam+\Delta$. 
For any other subderivation we conclude by the \ih\ and the fact
that each $\x:\D_i \in \Delta_i$  implies $\x:\D_i \in \Gam+\Del$.  
\qed

This lemma gives to system $\Mu$ a quantitative flavour.
In fact, define the {\it degree} of $(\Gam, \sig)$,
written $d(\Gam, \sig)$,  to be the sum of the cardinalities of all the multisets in $\Gamma$ and $\sigma$. The following property holds.
\begin{pty}\label{prop:times}
Let $\ap$ be an approximate normal form.
Let $\Pi \dem \Gam \der \ap:\sigma$. Then the number of 
typed positions of $\ap$ in $\Pi$ which correspond to 
both bound and free variables (\cf\ Def.~\ref{def:occ}) is
bounded by $d(\Gam, \sig)$.
\end{pty}

\proof Consider a (free or bound) variable $\x$ of $\s$, and let
$(\x:\multiset{\sig_i} \der \x:\sig_i)_{\iI}$ be all the axioms with
subject $\x$ in $\Pi$.  By the hygiene convention, $\x$ is either free
or bound in $\s$.  If $\x$ is free in $\s$, then
$\x:\multiset{\sig_i}_{\iI} \in \Gamma$, and since every axiom
corresponds to a typed position of $\x$ in $\Pi$, the proof is
given. If $\x$ is bound in $\s$, then there is a subderivation of
$\Pi$ ending with rule $(\arri)$, with premise $\Gam',
\x:\multiset{\sig_i}_{\iI} \der \ap': \tau$ and conclusion $\Gam' \der
\l \x.\ap':\multiset{\sig_i}_{\iI} \arrow \tau$. By the subformula
property (Lemma~\ref{lem:subformula}),
$\multiset{\sig_i}_{\iI} \arrow \tau$ is a subformula of either $\Gam$
or $\sig$, so the cardinality of $I$, which corresponds to the number
of bound typed positions of $\x$ in $\Pi$, is smaller or equal
to $d(\Gam,\sig)$.  \qed

This property has a very important corollary.
\begin{cor}
Given a pair $(\Gam, \sig)$, the number of approximate normal forms $\ap$ such that $\Pi\dem\Gam \der \ap:\sigma$ and $\ap=\Ap(\Pi)$ is finite.
\end{cor}
\proof By Property~\ref{prop:times}, if $\Pi \dem  \Gam \der \ap :\sig$, then
the number of typed positions of variables of  $\ap$ in $\Pi$ is bounded by
$d(\Gam,\sig)$. By definition of $\Ap(\Pi)$,
  $\Omega$ is the only untyped subterm of $\ap$, then $d(\Gam,\sig)$ gives
an upper bound for the number of {\it all} positions of variables 
of $\ap$. But the number of approximate normal forms with a bounded
number of variable positions  is finite, so the number of such
approximate normal forms $\ap$ is finite too.  \qed

\subsection{The inhabitation algorithm}
 \label{s:inhabitation}

An instance of the inhabitation problem is a pair  $(\Gamma,\sigma)$, and 
solving such an instance consists in deciding whether there exists a
term $\s$ such that $\Gamma\der \s: \sigma$  is derivable.

We find convenient to present inhabitation algorithms as deductive systems, proving  
judgements of the form $\ap\real\K(\Gamma,\sigma)$, whose intended meaning is that there exists a derivation  $\Pi\dem \Gamma\der \ap:\sigma$ such that $\ap=\Ap(\Pi)$. In this spirit, a  {\it run} of the algorithm on the input $(\Gamma,\sigma)$ is a proof of the judgement   $\ap\real\K(\Gamma,\sigma)$, for some approximate normal form $\ap$,
which is called the {\it output} of that run.
Hence, running the algorithm on
$(\Gamma,\sigma)$ corresponds to searching a proof
of $\ap\real\K(\Gamma,\sigma)$, for some (unknown) $\ap$.
As usual, when writing  $\ap\real\K(\Gamma,\sigma)$ we mean
that there exists a proof of that judgement.


When facing a problem $(\Gamma,\sigma)$, in case $\sigma=\D\arrow\tau$, there exist two possibilities. The first one consists in guessing
that the last rule used in  a derivation solving the problem is $(\arri)$, and trying to solve the problem $(\Gamma+\x:\D,\tau)$ for 
a fresh variable $\x$. The second  one consists in guessing that  the last rule is $(\arre)$, choosing a head variable among those in the domain of $\Gamma$ having 
a type of the form $\A_1\arrow\ldots\arrow \A_n\arrow\sigma$ (if any),  and trying to construct the arguments $\ap_1,...,\ap_n$ 
using the resources still avaliable in $\Gamma$. 

This gives the algorithm in  Figure~\ref{fig:inhabitation-algorithm-H}. 
The two alternatives described above are implemented by rule $(\Abs)$ and rule 
 $(\Head)$ respectively. Once the head variable has been chosen, by  $(\Head)$, 
rule  $(\Head_{>0})$ allows to construct its argument one by one, using 
rule $(\Union)$. All the branches of any given run of the algorithm stop
on a $(\Head_0)$ rule, or on a $(\Union)$ rule with $I=\emptyset$, or fail (no rule applies).

More precisely, the algorithm uses three forms of judgements: 
\begin{itemize}
\item $\ap\real\KI(\Gam, \A)$ meaning that the approximate normal form
$\ap$ solves the inhabitation problem $(\Gam, \A)$.

\item $\ap\real\K(\Gam, \sig)$,  meaning that the $\cal N$ approximate normal form
$\ap$ solves the inhabitation problem $(\Gam, \sig)$.

\item $\ap\real{\tt H}^{\x:\multiset{\tau}}(\Gamma,\sigma)$  meaning that the $\cal L$ approximate normal form
$\ap$, headed by a variable $\x$ of type $\tau$, solves the inhabitation problem $(\Gam, \sig)$.
\end{itemize}

\noindent We prove termination, soundness and completeness of the algorithm presented in Figure~\ref{fig:inhabitation-algorithm-H}. 
Completeness, in particular, is obtained  thanks to
a non-deterministic behaviour: given an environment
$\Gam$ and a type $\sigma$, different runs can be chosen, each one
constructing a  judgement of the form $\ap \real \K(\Gam, \sig)$.
By  collecting all such  possible runs 
we recover {\it all} the approximate normal forms $\ap$ such that there exists a derivation $\Pi\dem\Gam
\der \ap:\sig$, with $\ap=\Ap(\Pi)$.

\begin{figure}[!ht]
\[ \infer{\ap \real \K(\Gam +  \x:\A,   \tau)  \sep \sep 
   \x \notin \dom{\Gam} }
{ \l \x. \ap \real  \K(\Gam  ,  \A \arrow \tau)}\ (\Abs)
\]

\[ 
   \infer{ (\ap_i \real  \K( \Gam_i,  \sig_i))_{\iI}\sep \uparrow_{\iI}\ap_i }
         {\bigvee_{\iI} \ap_i \real   \KI(+_{ \iI} \Gam_i,  \multiset{\sig_i}_{\iI})}\ (\Union) 
         \]
\[ \infer{
 \Gam = \Gam_{1}+\Gam_{2} \sep \ap  \real  \LH{\x: \multiset{\A_1 \arrow \ldots \A_n\arrow 
\D \arrow \tau}}{(\Gam_{1}, \D \arrow \tau)} \sep \bp\real  \KI(\Gam_{2}, \D) \sep n\geq 0
}
 { \ap \bp  \real  \LH{\x: \multiset{\A_1 \arrow \ldots \A_n\arrow \D \arrow \tau}}{(\Gam, \tau)}}\ (\Head_{>0}) 
         \]
        
\[ \infer{ }
         { \x  \real  \LH{\x: \multiset{ \tau}}{(\es, \tau)}}\ (\Head_{0}) 
         \]
         
         \[
     \infer{\ap \real \LH{\x:\multiset{\A_1 \arrow \ldots \A_n \arrow \tau}}{(\Gam,\tau)}}
           {\ap \real \K(\Gam + \x:\multiset{\A_1 \arrow \ldots \A_n \arrow \tau},\tau)}  \ (\StartHead) 
         \]

\caption{The inhabitation algorithm for system $\Mu$}
\label{fig:inhabitation-algorithm-H}
\end{figure}

Some comments on the rules of the algorithm are in order. Rule
$(\Abs)$ looks for an ${\cal N}$ approximate normal form, when the
input type is an arrow. Rule
$(\Union)$ applies the approximation theorem \ref{th:approx}; notice
that in the particular case $I=\es$ it gives $\Omega\real
\KI(\es,\emul)$, where $\es$ denotes the environment having empty
domain. Rule $(\Head_{0})$ is self-explaining, rule $(\Head_{>0})$ is
based on the property that, if an approximate normal form is an
application $\ap\bp$, then $\ap$ must be an ${\cal L}$ approximate
normal form, and its type is a subtype of a type assigned to its head
variable. Rule $(\StartHead)$ selects a head variable and launches the search for a suitable 
 ${\cal L}$ approximate normal form.

\begin{exa}\label{Exe1}\mbox{}
\begin{enumerate}
\item\label{primo} Let $\Gamma=\es$ and $\sigma
  =\multiset{\multiset{\alpha} \to \alpha}\arrow \multiset{\alpha} \to
  \alpha$.  Given input $(\Gamma,\sigma)$, the algorithm
  succeeds with two different
  runs, generating respectively the following deduction trees:

(1.1)
\[
\begin{prooftree}
  \begin{prooftree}
    \begin{prooftree}
  \begin{prooftree}
    \begin{prooftree}
      \justifies{\x\real \LH{\x: \multiset{\multiset{\alpha} \to  \alpha}}{(\es,\multiset{\alpha} \to  \alpha)}}
      \using{(\Head_{0})}
    \end{prooftree} \sep
    \begin{prooftree}
      \begin{prooftree}
        \begin{prooftree}
          \justifies{\y \real \LH{\y: \multiset{\alpha}}{(\es,\alpha)}}
          \using{(\Head_{0})}
          \end{prooftree}        
        \justifies{\y \real \K(\y: \multiset{\alpha}, \alpha )}
        \using{(\StartHead)}
        \end{prooftree}
      \justifies{\y\real \KI(\y: \multiset{\alpha}, \multiset{\alpha})}
      \using{(\Union)}
      \end{prooftree}
      \justifies{\x\y \real \LH{\x: \multiset{\multiset{\alpha} \to  \alpha}}{(\y: \multiset{\alpha}, \alpha)}}
      \using{(\Head_{>0})}
    \end{prooftree}
    \justifies{\x\y \real \K(\x: \multiset{\multiset{\alpha} \to  \alpha}, \y: \multiset{\alpha}, \alpha)}
\using{(\StartHead)}
\end{prooftree}
\justifies{\l\y.\x\y \real  \K(\x: \multiset{\multiset{\alpha} \to  \alpha}, \multiset{\alpha} \to \alpha)}
\using{(\Abs)}
\end{prooftree}
\justifies{\l \x\y. \x\y \real \K( \es, \sig)}
\using{(\Abs)}
\end{prooftree} \]

(1.2)
\[
\infer{
\infer{
\infer{ }{\x \real \LH{\x: \multiset{\multiset{\alpha} \to  \alpha}}{(\es,\multiset{\alpha} \to \alpha)}} \ (\Head_{0})
}
{\x \real  \K(\x: \multiset{\multiset{\alpha} \to  \alpha}, \multiset{\alpha} \to \alpha)} \ (\StartHead)
}
{\l \x.\x \real \K(\es,\sig)} \ (\Abs)
\]
Remark that the type $\sig$ does not correspond to the simple type
$(\alpha \to \alpha)\arrow \alpha \to \alpha$, which represent the
data types of Church numerals and so has an infinite number of
inhabitants. In system $\Mu$ there is no common type for all the Church
numerals, since the numeral $\underline{n}$ has type
$\multiset{\underbrace{\multiset{\alpha} \to
    \alpha,...,\multiset{\alpha} \to \alpha}_n}\to
\multiset{\alpha}\to \alpha$ (among others, all of degree $n$).
\item  Let $\Gamma=\es$ and $\sigma =\multiset{\emul \to \alpha}\arrow \alpha$.
Given input $(\Gamma,\sigma)$, a succesful run of the algorithm is:
\[
\infer{
\infer{
\infer{
{\infer{ }{\x \real \LH{\x: \multiset{\emul \to \alpha}}{(\es, \emul \to \alpha)} }\ (\Head_{0)}) \sep
 \infer{ }{ \Omega \real \KI(\es, \emul)} \ (\Union)}
}
{\x \Omega \real \LH{\x: \multiset{\emul \to \alpha}}{(\es, \alpha)}} \ (\Head_{>0})}
{\x\Omega \real \K(\x: \multiset{\emul \to \alpha}, \alpha} \ (\StartHead)}
{\l \x .\x \Omega \real \K(\es, \sigma)} \ (\Abs)
\]

\item Given input $(\es, \multiset{\alpha_1}\to \alpha_2)$,
 where each $\alpha_i$ is a base type, then the unique possible
run consists in starting with  rule $(\Abs)$, then rule $(\StartHead)$, 
 then the algorithm stops since no other rule can be applied. Then 
$\K(\es, \multiset{\alpha_1}\to \alpha_2)$ is empty.
\end{enumerate}  
\end{exa} 
\noindent It follows from
Example~\ref{Exe1}.\ref{primo} that the algorithm is not an obvious extension of the classical
inhabitation algorithm for simple types~\cite{BenYellesPhd,hindley08},
and cannot be conservative with respect to it, since the two
algorithms take input types belonging to different grammars.

\begin{defi}
In order to show that the inhabitation algorithm terminates,
we define a {\it measure on types and environments}, as follows:

$ \begin{array}{lll@{\hspace{1cm}}lll}
   \ccount{\alpha} & = & 1 & \ccount{\multiset{\sig_i}_{\iI}} & = & \sum_{\iI} \ccount{\sig_i} +1 \\
   \ccount{\A \to \rho } & = & \ccount{\A}  + \ccount{\rho} + 1 
&\ccount{\Gam} & = & \sum_{\x\in \dom\Gamma}\ccount{\Gamma(\x)} \\
\end{array} $
\end{defi}
The measure is then extended to the judgements of the algorithm:
 
$\ccount{\K(\Gam, \rho)}=\ccount{\Gam}+\ccount{\rho}$

$\ccount{\KI(\Gam, \A)}=\ccount{\Gam}+\ccount{A}$

$\ccount{\LH{\x:\multiset{\A_1 \arrow \ldots \A_n \arrow  \tau}}{(\Gam, \tau)}}=\ccount{\Gam} +_{i=1 \ldots n} \ccount{\A_i} +n $

\begin{lem}[Termination]
\label{l:termination}
The inhabitation algorithm for system $\Mu$ terminates. 
\end{lem} 
\proof Given input $(\Gam,\sig)$, every run of the algorithm is a
deduction tree labelled with the rules
  of the algorithm (as in Example~\ref{Exe1}), where a node $n'$ is a
child of $n$ iff there exists an instance of a rule having $n$ as
conclusion and $n'$ as one of their premises. We associate to
every rule application the measure of its conclusion, and we will
prove that at every recursive call the measure decreases. The proof is
by induction on the deduction tree.

If the last applied rule is
$(\Head_{0})$, then the proof is trivial.
If the last rule is
$(\Abs)$ or $(\Union)$, then the proof is straightforward
by application of the \ih\ and the definition of the measure. 
So let us assume that  the last rule
is $(\StartHead)$, with premise $\ap \real \LH{\x:\multiset{\A_1
    \arrow \ldots \A_n \arrow \tau}}{(\Gam,\tau)}$ and conclusion $\ap
\real \K(\Gam + \x:\multiset{\A_1 \arrow \ldots \A_n \arrow
  \tau},\tau)$. Then the measure of the premise is $\ccount{\Gam}
+_{i=1 \ldots n} \ccount{\A_i} + n $, while that of the conclusion is
$ \ccount{\Gam} +_{i=1 \ldots n} \ccount{\A_i} + n + 2
\times\ccount{\tau} $, which is strictly bigger.
Let us assume that the last rule be $(\Head_{>0})$, with
premises $\ap \real \LH{\x: \multiset{\A_1 \arrow \ldots \A_n\arrow \D
    \arrow \tau}}{(\Gam_{1}, \D \arrow \tau)}$ and $\bp\real
\KI(\Gam_{2}, \D) $, and conclusion $ \ap \bp \real \LH{\x:
  \multiset{\A_1 \arrow \ldots \A_n\arrow \D \arrow \tau}}{(\Gam,
  \tau)}$.  The measure of the conclusion is $\ccount{\Gam}+_{i=1
  \ldots n} \ccount{\A_i} +\ccount{\D} + n +1$, while the measures of
the premises are respectively $\ccount{\Gam_{1}} + +_{i=1 \ldots n}
\ccount{\A_i} + n$ and $\ccount{\Gam_{2}} + \ccount{\D}$. Also taking
into account that either $\Gam_{1}$ or $\Gam_{2}$ can be empty, both
the premises have a measure strictly smaller than then  measure of the conclusion.

Hence any particular run of the algorithm terminates.

Since  any instance $\K(\Gam,\sigma)$ (resp.  $\KI(\Gam,\A)$ or   $\LH{\x:\sigma}{(\Gam,\tau)}$) gives  rise to a finite number of possible runs,  as in Example~\ref{Exe1}, we conclude. \qed




The definition below gives the intended meaning of the three components of the inhabitation algorithm:

\begin{defi}\label{Def:sem}\mbox{}
\begin{itemize}
\item $\interp{\K(\Gam, \sig)} = \set{\ap \mid \exists \Pi.\ \Pi  \dem \Gam \vdash \ap:\sig\mbox{ and } \ap=\Ap(\Pi) }$, 
\item $\interp{\KI(\Gam, \A)} = \set{ \ap \mid \exists \Pi.\ \Pi  \dem \Gam \vdash \ap:\A\mbox{ and } \ap=\Ap(\Pi)  }$, 
\item $\interp{\Hn \x {\A_1\arrow\ldots\arrow A_n\arrow} \tau \Gam} = \set{ \x\ap_1\ldots\ap_n \mid 
\exists \Pi.\  \Pi  \dem  \Gam + \x:\mult{\A_1\arrow\ldots\arrow \A_n\arrow \tau}  \vdash \x\ap_1...\ap_n :\tau  \mbox{, }  
 \x\ap_1...\ap_n=\Ap(\Pi)  \mbox{, }   \Gam=+_{i=1,\ldots,n}\Gam_i \mbox{ and }  \Gam_i\der \ap_i:\A_i}$.
\end{itemize}
\end{defi}

Remark that $\interp{\Hnp \x {\A_1\arrow\ldots\arrow \A_n\arrow \tau} \Gam \tau}
  \subseteq \interp{\K(\Gam +\x:\multiset{\A_1\arrow\ldots\arrow \A_n\arrow \tau}, \tau)}$.  
Moreover, if $\ap \in \interp{\K(\Gam, \sig)}$ and $\ap = \x\ap_1...\ap_n$, then 
$\ap \in \interp{\Hnp \x {T_\tau} {\Gam'} \tau}$
for some type $T_\tau = \A_1\arrow\ldots\arrow \A_n\arrow \tau$ such that $\Gam = \Gam' + \x:\multiset{T_\tau}$. 

 Soundness and completeness of the
inhabitation algorithm follow from the following Lemma, relating
typing derivations in system $\Mu$ and runs of the algorithm:

\begin{lem}\label{Lem:main}
Let $\ap$ be an approximate normal form, $\Gamma$ a typing environment
and $\sig$ a type. Then $\ap\real\K(\Gamma,\sigma) \Leftrightarrow
\ap \in \interp{\K(\Gamma,\sigma)}$.
\end{lem}

\proof

As for ($\Rightarrow$), we prove the following:

\begin{enumerate}[label=\alph*)]
\item
$\ap\real\K(\Gamma,\sigma)\Rightarrow \ap \in \interp{\K(\Gamma,\sigma)}$.
\item
$\ap\real\KI(\Gamma,\A)\Rightarrow\ap \in \interp{\KI(\Gamma,\A)}$. 
\item
$\ap\real  \Hn \x {\A_1\arrow\ldots\arrow A_n\arrow} \tau \Gam                           \Rightarrow
\ap \in \interp{ \Hn \x {\A_1\arrow\ldots\arrow \A_n\arrow} \tau \Gam     }$.

\end{enumerate} 

These three properties are proved by mutual induction on 
{the  definitions of
the jugdments $\ap\real\K(\Gamma,\sigma)$, $\ap\real\KI(\Gamma,\A)$ and $\ap\real  \Hn \x {\A_1\arrow\ldots\arrow A_n\arrow} \tau \Gam $}. 

\begin{enumerate}[label=\alph*)]
\item There are two cases:

\begin{itemize}

\item If $\l\x.\ap\real \K(\Gamma,\A\arrow \tau)$ follows from 
 $\ap\real \K(\Gamma+(\x :\A),\tau)$ by $(\Abs)$,  then we conclude by the \ih\ (a)
and by an application of rule ($\arri$). 

\item If $\ap\real \K(\Gamma+(\x :\multiset{\A_1\arrow\ldots\arrow
  \A_n\arrow \tau}),\tau)$ follows from $\ap \real \Hn \x
  {\A_1\arrow\ldots\arrow A_n\arrow} \tau \Gam$, then we conclude
  immediately by the \ih\ (c).  Note that in this case $\ap$ is of the
  form $\x\ap_{1}...\ap_{n}$.
\end{itemize}

\item[b)] If $\bu_ {\iI}\ap_i\real\KI(+_{\iI}\Gamma_i,\multiset{\sigma_i}_{\iI})$ follows from
$(\ap_i\real\K(\Gam_i,\sigma_i))_{\iI}$ and $\uparrow_{\iI}\ap_i$ by $(\Union)$, then by 
the \ih\ (a), for all $\iI$ there exists 
 $(\Pi_i\dem\Gam_i\der \ap_i:\sigma_i)_{\iI}$ such that 
$\ap_i=\Ap(\Pi_i)$. By Proposition \ref{Prop:monoton}.\ref{monone}, for all $\iI$ there exists   $\Pi'_i\dem\Gam_i\der\bu_{\iI}\ap_i:\sigma_i$, with $\Ap(\Pi'_i)=\Ap(\Pi_i)$.
By rule $(\many)$ we obtain $\Pi_\many\dem +_{\iI}\Gamma_i \der\bu_ {\iI}\ap_i:\multiset{\sigma_i}_{\iI}$.
We conclude by observing that $\bu_ {\iI}\ap_i=\bu_ {\iI}\Ap(\Pi_i)= \bu_ {\iI}\Ap(\Pi'_i)=  \Ap(\Pi_\many)$.

\item[c)] There are two cases:

\begin{itemize}
\item If $\x\real  \Hn \x {} \tau \Gam  $ is an axiom, then  $\x= \Ap(\Pi)$
where $\Pi$ is the unique proof  of 
$\x:[\tau]  \vdash \x :\tau$. 

\item If  $\ap\bp \real  \Hn \x {\A_1\arrow\ldots\arrow A_n\arrow} \tau {\Gam+\Delta}$ follows from 
$\ap \real  \Hn \x {\A_1\arrow\ldots\arrow \A_{n-1}\arrow} {\A_n\arrow \tau} \Gam$ and 
$\bp\real \KI(\Delta, \A_n)$, then:

\begin{itemize}
\item
by the \ih (c), $\ap=\x\ap_1...\ap_{n-1}$, there exists $\Pi_1\dem \Gam + \x:[\A_1\arrow\ldots\arrow A_n\arrow\tau] \der \ap : \A_n\arrow \tau$
such that  $\ap=\Ap(\Pi_1)$.

\item
by the  \ih (b) there exists $\Pi_2\dem \Delta \der \bp:\A_n$  such that  $\bp=\Ap(\Pi_2)$.
\end{itemize}

By using Rule  $(\arre)$, we obtain a proof $\Pi\dem \Gam+\Delta + \x:[\A_1\arrow\ldots\arrow A_n\arrow\tau] \der \x\ap_1...\ap_{n-1}\bp:\tau$, with 
$\Ap(\Pi)=\x\ap_1...\ap_{n-1}\bp=\ap\bp$, and we are done.
\end{itemize}

\end{enumerate}

\noindent Concerning ($\Leftarrow$), we prove the following:

\begin{enumerate}[label=\alph*)]
\item
$\ap\real\K(\Gamma,\sigma)\Leftarrow \ap \in \interp{\K(\Gamma,\sigma)}$.
\item 
$\ap\real\KI(\Gamma,\A)\Leftarrow\ap \in \interp{\KI(\Gamma,\A)}$. 
\item
$\ap\real  \Hn \x {\A_1\arrow\ldots\arrow A_n\arrow} \tau \Gam                           \Leftarrow
\ap \in \interp{ \Hn \x {\A_1\arrow\ldots\arrow A_n\arrow} \tau \Gam     }$.

\end{enumerate}

We proceed by induction on typing derivations $\Pi\dem\Gam\der\ap:\sigma$  or $\Pi\dem\Gam\der\ap:\A$
such that $\ap=\Ap(\Pi)$. 
For such derivations, we show that $\ap$ is generated by a suitable run of the 
inhabitation algorithm.

\begin{itemize}
\item Case $(\axvar)$. Then there is a type derivation of the following form: 
$$\infer{\mbox{}}{\Pi\dem\x:\multiset{\rho} \der \x:\rho}$$
In this case $\Ap(\Pi)=\x$ belongs to 
$\interp{\K(\x:\multiset{\rho},\rho)}$ and to $\interp{ \Hn \x {} \rho \emptyset}$,
so we need to show a) and c) respectively.

By rule $(\Head_{0})$ we have $\x\real \Hn \x {} \rho \emptyset$, then
by rule $(\StartHead)$ we also conclude
$\x\real\K(\x:\multiset{\rho},\rho)$.

\item  Case $(\arri)$.  Then there is a type derivation of the following form: 

$$\infer
{\Pi'\dem\Gam\der \bp:\tau }
{\Pi\dem\Gam \sm  \x \der \l \x. \bp: \Gamma(\x)\arrow \tau}\ (\arri)$$

with  $\l \x. \bp=\Ap(\Pi)$,  and hence  $\bp=\Ap(\Pi')$.
In this case,  $\l \x. \bp \in \interp{\K(\Gam \sm  \x, \Gamma(\x)\arrow \tau)}$
(and thus $\bp\in \interp{\K(\Gam, \tau)}$), 
so that we need to show a). 
By the \ih (a)  we have that $\bp\real \K(\Gam, \tau)$, and we conclude that  $\l \x. \bp \real \K(\Gam \sm  \x, \Gamma(\x)\arrow \tau))$
by rule  $(\Abs)$.
\item Case $(\arre)$. Then  $\ap=\x\bp_{1}...\bp_{n}$ for some $n>0$ and the typing
derivation $\Pi$ has the following form:
$$\begin{prooftree}
\Pi^{1} \dem \Gam_{1} \vdash \x\bp_{1}...\bp_{n-1}:\A_n \arrow \tau \sep
\Pi^{2}\dem \Gam_{2}\der \bp_{n}:\A_n
\justifies{\Gam \vdash \x\bp_{1}...\bp_{n}: \tau}
\using{(\arre)}
\end{prooftree}$$
where $\Gam = \Gam_1 + \Gam_2$. We analyse two cases:
\begin{itemize}
\item $\ap \in \interp{ \Hnp \x {T_\tau} {\Gam'} \tau    }$,
where $\Gam = \Gam' + \x:\multiset{T_\tau}$. 
Then $\Gam_1 = \Gam'_1 + \x:\multiset{T_\tau}$, and
$\Gam' =  \Gam'_1 + \Gam_2$. Moreover, 
  $ \x\bp_{1}...\bp_{n-1} = \Ap(\Pi^{1})$ and $\bp_{n}= \Ap(\Pi^{2})$.  By the \ih (c),
  $\x\bp_{1}...\bp_{n-1} \real \Hnp \x {T_\tau }  {\Gam'_{1}} {\A_n\arrow \tau}$ and, by the \ih (b)
  $\bp_{n}\real\KI(\Gamma_2,\A_n)$, so we get $\ap \real \Hnp \x {T_\tau}  {\Gam'} {\tau}$  by rule
  $(\Head_{>0})$.
 
\item  $\ap \in \interp{ \K(\Gam, \tau)}$. Then 
$\ap \in \interp{ \Hnp \x {T_\tau} {\Gam'} \tau     }$ 
for some type $T_\tau$
such that $\Gam = \Gam' + \x:\multiset{T_\tau}$ as remarked after Definition ~\ref{Def:sem}. 
Then we conclude $\ap \real \Hnp \x {T_\tau} {\Gam'} \tau$  by the previous point and 
$\ap \real \K(\Gam, \tau)$
by rule $(\StartHead)$. 
\end{itemize}

\item Case $(\many)$: If $\ap \in \interp{\KI(\Gamma,\A)}$, then $\Pi  \dem \Gam \der \ap:\A$, so we have a derivation of the following form:
$$\infer{(\Pi_i\dem\Delta_{i} \der \ap:\sig_i)_{\iI}} {\Pi\dem +_{\iI} \Delta_{i} \der \ap:\multiset{\sig_{i}}_{\iI}}(\many)$$
with $\Gam = +_{\iI} \Delta_{i}$, $\A= \multiset{\sig_{i}}_{\iI}$ and $\ap=\Ap(\Pi)=\bigvee_{\iI} \Ap(\Pi_i)$. In this case we need to prove  $\ap \real \KI(+_{\iI} \Delta_{i},\multiset{\sig_{i}}_{\iI} )$.
By
Proposition~\ref{Prop:monoton}.\ref{montwo} we obtain, for all $\iI$,
a proof $\Pi'_i\dem\Gam_i\der \Ap(\Pi_i):\sigma_i$ such that
$\Ap(\Pi'_i)=\Ap(\Pi_i)$. By the \ih (a) we have that for all $\iI$, $\Ap(\Pi_i)\real
\K(\Delta_i, \sigma_i)$, hence we conclude $\ap=\bigvee_{\iI}\Ap(\Pi_i)\real \KI(+_{\iI}
\Delta_{i},\multiset{\sig_{i}}_{\iI} )$ by rule $(\Union)$.
\end{itemize}

\qed

\begin{thm} [Soundness and Completeness of the inhabitation algorithm for $\Mu$] \mbox{}
\label{t:soundness-completeness-for-H}
\begin{enumerate} 
\item \label{t:soundness-H}
If $\ap \real \K(\Gam, \sig)$  then, for all $\s$ such that $\ap\leq\s$,  $\Gam  \der \s:\sig$.
\item \label{t:completeness-H}
If $\Pi\dem\Gam \der \s:\sig$ then there exists  a $\beta$-reduct $t'$ of $t$ and a derivation $\Pi'\demnf \Gam \der \s':\sig$ such that
  $\Ap(\Pi') \real \K(\Gam,\sig)$.
\end{enumerate}
\end{thm}
\proof
Soundness: if $\ap \real \K(\Gam, \sig)$ then by Lemma~\ref{Lem:main} ($\Rightarrow$) we have that $\Gam\der \ap:\sig$.
Then, by Proposition~\ref{Prop:monoton}.\ref{montwo} we get   $\Gam  \der \s:\sig$ for all  $\ap\leq\s$.

Completeness: if $\Pi\dem\Gam \der \s:\sig$ then by Theorem~\ref{thr::char}.\ref{th::red1}, there exists
$\Pi'\demnf\Gam \der \s':\sig$, with $\s\arrow^*_\beta\s'$. By Proposition~\ref{Prop:monoton}.\ref{montwo}, 
there exists $\Pi''\dem \Gamma\der \Ap(\Pi'):\sigma$, and $\Ap(\Pi'')=\Ap(\Pi')$. 
We conclude that   $\Ap(\Pi') \real \K(\Gam,\sig)$ by  Lemma~\ref{Lem:main}
($\Leftarrow$). \qed


  The main difference between our inhabitation algorithm for
  system $\Mu$ and the classical inhabitation algorithm for simple
  types~\cite{BenYellesPhd,hindley08} lies in the use of
  approximants. Besides that, our algorithm is peculiar in the
  fact that the head variables' arguments are constructed one by one,
  in such a way that the runs of the algorithm mirror exactly the
  typing derivations of system $\Mu$. As a matter of fact, a less
  fine-grained version of the algorithm is also possible, where all
  the arguments of the head variable are constructed at once (see
  Figure~\ref{fig:inhabitation-algorithm-simple}).  Our step-by-step
  version is a simplified version of the algorithm presented in
  \cite{DBLP:conf/ifipTCS/BucciarelliKR14}. It is worth noticing that
  the simplification has a price: both the systems in
    Fig.\ref{fig:inhabitation-algorithm-H} and
    \ref{fig:inhabitation-algorithm-simple} are not linear, 
because of the rules $(\Head_{\mathtt b})$ and $(\Head_{>0})$: the type $\tau$ occurs in the conclusion of those
  rules both as the target type and as a subtype of the head variable's
  type.

\begin{figure}[!ht]
\[ \infer{\ap \real \K(\Gam +  (\x:\A) ,   \tau)  \sep \sep 
   \x \notin \dom{\Gam} }
{ \l \x. \ap \real  \K(\Gam  ,  \A \arrow \tau)}\ (\Abs)
\]

\[ 
   \infer{ (\ap_i \real  \K( \Gam_i,  \sig_i))_{\iI}\sep \uparrow_{\iI}\ap_i }
         {\bigvee_{\iI} \ap_i \real   \KI(+_{ \iI} \Gam_i,  \multiset{\sig_i}_{\iI})}\ (\Union) 
\]

\[ \infer{ \Gam = +_{i=1\ldots n} \Gam_i \sep
           (\bp_i \real  \KI(\Gam_i, \A_i))_{i=1 \ldots n} }
         { \x \bp_1 \ldots \bp_n  \real  \K(\Gam +  (\x: \multiset{\A_1 \arrow \ldots \A_n \arrow \tau}), \tau)}\ (\Head_{\mathtt b}) \]
\caption{The basic inhabitation algorithm for system $\Mu$}
\label{fig:inhabitation-algorithm-simple}
\end{figure}

\subsection{Breaking relevance}

Among the non-idempotent intersection type systems, $\Mu$ has certainly a  particular status, since it 
induces a denotational model of the $\lambda$-calculus, based on relational semantics ~\cite{PPR17}. Thus 
our investigation of the inhabitation problem for 
non-idempotent intersection types systems started with $\Mu$.
The system $\Mu$ is relevant,
and it is reasonable to suppose that relevance plays a role in the decidability of the inhabitation problem, 
as it allows for a fine control of resource management.
Hence, breaking relevance is our next step in the investigation  pointed out above. 
Adding weakening to the type system 
 is  semantically unsound, in this framework,  in the sense that the extended system does not induce  a $\lambda$-model.
Nevertheless it enjoys several interesting properties, and in particular the approximation theorem continues to hold.

It turns out that adding weakening to the system $\Mu$ does not break the decidability of the type inhabitation problem.
The system  $\Mu_w$, presented in Figure~\ref{fig:W},
is obtained by relaxing the axiom  $(\axvar)$ of  system $\Mu$ to the more general axiom  $(\axvarw)$, 
where the context is possibly over-defined, meaning that  it may assign several different  types both to  the subject $\x$
and to other variables.
\begin{figure}[!ht]
\begin{center}
$\begin{array}{c}
\infer{1 \leq i \leq n}{\Gamma, \x:\multiset{\rho_1,...,\rho_n} \der \x:\rho_i}\ (\axvarw)
\quad \quad 
\infer
{\Gam\der \s:\tau }
{\Gam \sm  \x \der \l \x. \s: \Gamma(\x)\arrow \tau}\ (\arri)\\ \\
\infer
{\Gam\der \s: \A \arrow\tau  \sep 
\Delta\der \uu:\A }
{\Gam +\Delta\der \s\uu: \tau}\ (\arre)\qquad
\infer{(\Delta_{i} \der \s:\sig_i)_{\iI}} {+_{\iI} \Delta_{i} \der \s:\multiset{\sigma_i}_{\iI}}(\many)\\  \\
\end{array}$
\caption{The type assignment system $\Mu_w$}\label{fig:W}
\end{center}
\end{figure}

It is easy to see that the usual weakening rule:
\[\infer{\Gam \der \s:\sigma}{ \Gam + \Delta \der \s:\sigma} \ (\mathtt{w})\]
is admissible in system  $\Mu_w$.
Moreover,  $\Mu_w$  is still syntax directed, and thus in particular $\Pi
\dem \Gam \der \s:\sigma$ and $\Pi' \dem \Gam + \Delta \der \s:\sigma$
imply $\meas (\Pi) = \meas (\Pi')$.  Note that, while in the
idempotent case the presence of weakening makes the multiplicative and
the additive versions equivalent, this is no longer true in the
non-idempotent case. In fact, in order to preserve decidability of
inhabitation, we need to preserve the quantitative flavour of the
system. Indeed, for $\Mu_w$, Property~\ref{prop:times} still holds.
Clearly, in case of an additive definition of rule $(\arre)$,  
like in Figure~\ref{fig:C},
that property does not hold.  It is quite easy to check that
system   $\Mu_w$ enjoys  subject reduction, subject expansion and that 
it characterizes head-normal
forms.  
The algorithm proving the decidability of the type inhabitation problem for $\Mu_w$
can be easily obtained from the one in
Figure~\ref{fig:inhabitation-algorithm-H} by
changing the rule $(\Head_{0})$ as follows:
\[ \infer{ }
         { \x  \real  \LH{\x: \multiset{ \tau}}{(\Gam, \tau)}}\ (\Head_{0}) 
         \] 
This new rule takes into account the fact that, in presence of
weakening, not all the resources in the type context need to be
consumed during a derivation. So, the algorithm may stop and produce a
variable even if the context has not been fully
  consumed, assuming that the unused resources are created by
weakening, and can be discarded.  The same soundness and completness
arguments of Theorem~\ref{t:soundness-completeness-for-H} apply, with
the obvious changes.  We can conclude that the undecidability of the
inhabitation problem for systems like those in Figures~\ref{fig:C}
and~\ref{fig:S} is due to the idempotency of the intersection, and not
to the lack of relevance.

\section{Systems characterizing strong normalization}\label{section:sn}

\subsection{Disallowing untyped formal parameters}\label{section:sn1}
\label{section:fa}

Apparently the easiest way to modify the system $\Mu$ in order to
restrict the class of typable terms to the set of strongly normalizing
terms is to forbid untyped subterms inside typed terms. 

The technical way to obtain this behaviour is to mimic what happens in
the simply type assignment system, where, if we want to abstract a
term with respect to a variable $\x$ which is not in the domain of the
context, we just guess a type for it. Consequently, the empty multiset
is no longer  a multi-type, and the grammar of types becomes:
\[ \begin{array}{llllll}
\sigma, \tau, \rho & ::= & \alpha \mid \A \arrow \tau & (\mbox{types})  \\  
\A  & ::= & \multiset{\sigma_{i}}_{\iI}\ (I \neq \es)  &  (\mbox{multiset types}) 
\end{array} \] 
The resulting type system $\Mu_{e}$ is presented in Figure~\ref{fig:MVide}.
\begin{figure}[!ht]
\begin{center}
$\begin{array}{c}
\infer{\mbox{}}{\x:\multiset{\rho} \der \x:\rho }\ (\axvar) \\ \\
\infer
{\Gam \der \s:\tau\ \sep  \x \in \dom{\Gam}}
{\Gam \sm \x \der \l \x. \s:  \Gam(\x) \arrow \tau}\ (\arrine)
\hspace{1cm}
\infer
{\Gam \der \s:\tau\  \sep \x \notin \dom{\Gam} }
{\Gam \der \l \x. \s:  \multiset{\sig} \arrow \tau}\ (\arrie) \\ \\
\infer
{\Gam\der \s: \A\arrow\tau \sep \Delta\der \uu:\A}
{\Gam + \Delta\der \s\uu: \tau}\ (\arre) \quad \quad 
\infer{(\Delta_{i} \der \s:\sig_i)_{\iI} \quad I \not= \es} {+_{\iI} \Delta_{i} \der \s:\multiset{\sigma_i}_{\iI} }(\many)\\ \\
\end{array}$
\caption{The type assignment system $\Mu_{e}$} 
\label{fig:MVide}
\end{center}
\end{figure}

The type $\sig$ in  rule $(\arrie)$ is arbitrary, so that it is chosen  non-deterministically.
Remark also that 
untyped subterms are no longer allowed inside typed terms, since the $(\arre)$ rule
always impose a functional type with non-empty multiset type on its left hand side. 
System $\Mu_{e}$  does not enjoys  subject reduction nor
subject expansion. In fact $\x:\multiset{\sigma} \der (\lambda \y.\iid)\x: \multiset{\tau}
\arrow \tau$, while $\x:\sigma \not\der \iid: \multiset{\tau} \arrow
\tau$. In contrast, in the simple type case, the subject reduction  is verified 
since the system allows weakening.

So, in order to restore subject reduction, we add 
weakening to  system $\Mu_{e}$. The resulting system  $\Mu_{e,w}$ is presented in
Figure~\ref{fig:MWVide}. 

\begin{figure}[!ht]
\begin{center}
$\begin{array}{c}
\infer{ 1 \leq i \leq n}{\Gamma, \x:\multiset{\rho_1,...,\rho_n} \der \x:\rho_i}\ (\axvarw)
\quad \quad 
\infer
{\Gam \der \s:\tau \sep \x \in \dom{\Gam}}
{\Gam \sm \x \der \l \x. \s: \Gam(\x)\arrow \tau}\ (\arri)\\ \\
\infer
{\Gam\der \s: \A \arrow\tau \sep 
 \Delta\der \uu:\A }
{\Gam + \Delta\der \s\uu: \tau}\ (\arre)
\qquad
\infer{(\Delta_{i} \der \s:\sig_i)_{\iI}} {+_{\iI} \Delta_{i} \der \s:\multiset{\sigma_i}_{\iI}}(\many)
\end{array}$
\caption{The type assignment system $\Mu_{e,w}$} 
\label{fig:MWVide}
\end{center}
\end{figure}

The weakening is only introduced in the axioms,
as for system $\Mu_w$, making  the rule:
\[\infer{\Gam \der \s:\sigma}{ \Gam + \Delta \der \s:\sigma} \ (\mathtt{w})\]
admissible.
The resulting system is still syntax directed, so that $\Pi \dem \Gam
\der \s:\sigma$ and $\Pi' \dem \Gam + \Delta \der \s:\sigma$ imply
$\meas (\Pi) = \meas (\Pi')$. Note that, as for system $\Mu_w$, the
rule $(\arre)$ is multiplicative, thus preserving the fact that
$d(\Gam,\sig)$ is an upper bound to the number of variable
positions
in any subject of a derivation with context $\Gam$ and type
$\sig$. Let us remark that, in contrast to system $\Mu_{e}$, 
 just one rule introducing the arrow is needed, since the presence of the
weakening ensures that, if $\Pi \dem \Gamma \der \s:\sig$, then
there is always $\Pi' \dem \Gamma' \der \s:\sig$,
where $\Gam'$ extends $\Gam$ and $\x \in \dom{\Gam'}$.

It is important to stress the fact that,  in presence of
weakening, the types assigned to a term only  depend on  the axioms
giving types to its free variables. In fact, if $\x$ does not occur in
the subject of a derivation, no axiom with subject $\x$ is used in the
derivation itself. 
Formally:
\begin{pty}\label{prop:fv} \mbox{} 
\begin{enumerate}
\item $\Pi \dem \Gamma, \x: \A\der_{\Mu_{e,w}} \s:\sigma$ and $\x\not\in \fv{\s}$ imply 
that  there are no axioms in $\Pi$ 
with subject $\x$, so $\x:\A$ has necessarily  been    introduced by 
weakening some other axiom. 
\item If $\Pi \dem\Gamma, \x: \A\der_{\Mu_{e,w}} \s:\sigma$ and $\x\not\in \fv{\s}$,
then  $\Pi'\dem \Gam \der_{\Mu_{e,w}} \s:\sigma$. 
\end{enumerate}
\end{pty}
The system enjoys the following good properties.
\begin{thm} \mbox{}
  \label{prop:ew}
\begin{enumerate}
\item (Subject reduction) $\Gamma \der_{\Mu_{e,w} }\s:\sigma$ and $\s \redb \uu$ imply $\Gam \der_{\Mu_{e,w}} \uu:\sigma$.
\item (Typed subject expansion) Let $\ccontext$ be a context.
Then $\Gamma \der_{\Mu_{e,w}} \ccontext[\vv\isubs{\w}{\x}]:\sigma$ and 
$\Delta \der \w:\A$  imply  
$\Gam' \der_{\Mu_{e,w}} \ccontext[(\l\x.\vv)\w]:\sigma$, for some $\Gam'$.
\item (Strong normalization) $\Gamma \der_{\Mu_{e,w}} \s: \sigma$ iff $\s$ is strongly normalizing.
\end{enumerate}
\end{thm}
\proof \mbox{}
  
\begin{enumerate}
\item We will prove something more, namely that $\Pi\dem \Gamma \der
  \s:\sigma$ and $\s \redb \uu$ imply $\Pi' \dem\Gam \der \uu:\sigma$,
  where $\meas(\Pi') < \meas (\Pi)$. Indeed, $\s \redb \uu$ means $\s =
  \ccontext[(\l \x.\vv)\w]$ and
  $\uu=\ccontext[\vv\isubs{\w}{\x}]$ for some context $\ccontext$. The proof is by induction 
  on
  $\ccontext$. For the base case $\ccontext = \square$, it is
  sufficient to consider an erasure reduction, \ie\ the situation
  where $\x \not\in \fv{\vv}$, being the not erasing case
  already proved \eg\ in~\cite{BKV17}. The
  derivation $\Pi$ for $\Gamma \der (\l \x.\vv)\w:\sigma$ is of the shape

\[ \begin{prooftree}
   \begin{prooftree}
   \Pi' \dem\Gam \der \vv:\sigma
   \justifies{\Gam \sm \x \der \l\x.\vv: \Gam(\x) \arrow \sigma}
   \using{(\arri)}
   \end{prooftree}
   \sep
   \Delta\der \w:\A
   \justifies{(\Gam \sm \x) + \Delta\der (\l \x.\vv)\w: \sigma}
   \using{(\arre)}
   \end{prooftree} \]

where $\x\not\in \fv{\vv}$ implies by Proposition~\ref{prop:fv}
that there is no axiom rule with subject $\x$.
So from $\Pi'\dem\Gam \der \vv:\sigma$, and the admissibility of the weakening rule,
there is a derivation $\Pi''\der\Gam + \Delta\der \vv:\sigma$. Note that $\meas (\Pi'') =
\meas (\Pi')$.
Since $\vv\isubs{\w}{\x} = \vv$, then $\Pi''$ is the desired derivation, and the case is proved, being $\meas (\Pi'') < \meas (\Pi)$. The other cases come easily by induction.

\item By induction on $\ccontext$. Let us consider the base case,
  \ie\ $\ccontext=\square$. Also in this point, we will consider only
  the case when $\x \not\in \fv{\vv}$, so that $\vv \isubs{\w}{\x} = \vv$, being the not erasing case   treated in~\cite{BKV17}.  Then,
  Property~\ref{prop:fv} gives  $\Gamma\sm \x \der \vv:\sigma$. By hypothesis
  $\Delta \der \w:\tau$, for some $\Delta$ and $\tau$, so by the admissibility of weakening
  we can build a derivation $\Gamma\sm \x + \x: \multiset{\tau} \der
  \vv:\sigma$, and then, by applying rule $(\arri)$, we get $\Gamma\sm \x \der
  \l\x.\s:\multiset{\tau} \arrow \sigma$. By rule  $(\arre)$ we obtain
  $\Gamma\sm \x + \Delta \der (\l\x.\vv)\w: \sigma$ as desired. The inductive case
  is straightforward.
\item   Full details can be found in~\cite{BKV17}. \qed
\end{enumerate}

\noindent Note that the proof of the third point of the previous theorem
guarantees that all normal forms are typed in the system, since the
set of normal forms is included in the  set of strongly-normalizing terms. 

The inhabitation problem for system $\Mu_{e,w}$ is decidable, 
the corresponding algorithm is given in the Figure~\ref{fig:sn-algorithm}.

           The
reader may remark that there are two rules which have been changed with
respect to the algorithm for system $\Mu$ in
Figure~\ref{fig:inhabitation-algorithm-H}: rule $(\Head)$ and rule
$(\Union)$. Rule $(\Head)$ is changed in the same way
we did for system
$\Mu_w$, and so the same considerations hold.
Rule $(\Union)$ in Figure~\ref{fig:inhabitation-algorithm-H} was building the set of all the approximate normal forms $\ap=\bu_{\iI} \ap_i$ such that  $\Gam =  +_{\iI} \Gam_i$,
$\ap_i\real\K(\Gam_i,\sigma_i)$ for all $i\in I$, and $\uparrow_{\iI}\ap_i$. In system $\Mu_{e,w}$ the situation is easier, since normal forms are pairwise unbounded (\ie\ there is no common upper bound). So the set $\KI(+_{\iI} \Gam_i, \multiset{\sig_i}_{\iI})$ now contains the unique normal form $\s$ such that 
$\Gam_i \der \s: \sig_i$, for all $\iI$.  
The algorithm is sound and complete, as the next theorem shows.
\begin{thm} [Soundness and Completeness for $\Mu_{e,w}$]
  \label{t:soundness-completeness-for-Hew}
  \mbox{}
\begin{enumerate} 
\item \label{t:soundness-Hwe}
If $\s\real \K(\Gam, \sig)$  then $\Gam  \der_{\Mu_{e,w}} \s:\sig$.

\item \label{t:completeness-Hew}
If $\Gam \der_{\Mu_{e,w}} \s:\sig$ then $\s'\real \K(\Gam, \sig)$, where $\s \redbs \s'$ and  $\s'$ is in normal form.

\end{enumerate}
\end{thm}
The proof is similar to the corresponding proof for the system $\Mu$, but simpler,
because of the observations made  before.

\begin{figure}[!ht]

\[ \infer{\s \real \K(\Gam +  \x:\A,   \tau)  \sep \sep 
   \x \notin \dom{\Gam} }
{ \l \x. \s \real  \K(\Gam  ,  \A \arrow \tau)}\ (\Abs)
\]

\[ 
   \infer{ (\s \real  \K( \Gam_i,  \sig_i))_{\iI} }
         {\s \real   \KI(+_{ \iI} \Gam_i,  \multiset{\sig_i}_{\iI})}\ (\Union) 
         \]
\[ \infer{
 \Gam = \Gam_{1}+\Gam_{2} \sep \s  \real  \LH{\x: \multiset{\A_1 \arrow \ldots \A_n\arrow 
\D \arrow \tau}}{(\Gam_{1}, \D \arrow \tau)} \sep \uu\real  \KI(\Gam_{2}, \D) \sep n\geq 0
}
 { \s \uu  \real  \LH{\x: \multiset{\A_1 \arrow \ldots \A_n\arrow \D \arrow \tau}}{(\Gam, \tau)}}\ (\Head_{>0}) 
         \]
        
\[ \infer{ }
         { \x  \real  \LH{\x: \multiset{ \tau}}{(\Gam, \tau)}}\ (\Head_{0}) 
         \]
         
         \[
     \infer{\s \real \LH{\x:\multiset{\A_1 \arrow \ldots \A_n \arrow \tau}}{(\Gam,\tau)}}
           {\s \real \K(\Gam + \x:\multiset{\A_1 \arrow \ldots \A_n \arrow \tau},\tau)}  \ (\StartHead) 
         \]

\caption{The inhabitation algorithm for system $\Mu_{e,w}$}
\label{fig:sn-algorithm}

\end{figure}

We conclude this section by showing that the inhabitation problem for $\Mu_{e}$ reduces to the one for $\Mu_{e,w}$. 
\begin{lem}\label{lemma:H_w_increasing}    Let $(\Gamma,\tau)$ be an instance of the inhabitation problem
having a solution in system $\Mu_e$. Then 
 $(\Gamma+\x:\multiset{\sigma},\tau)$ has also a solution in  $\Mu_e$,
 for each variable $\x$ and type $\sigma$.
\end{lem}
\proof
A suitable type derivation is the following:
$$
\infer
{\infer
{\infer
{\vdots}{\tyj{\s}{\Gamma}{\tau}\sep\z\not\in\fv{\s}}
}
{\tyj{\l \z.\s}{\Gamma}{\mult{\sigma}\to\tau}}\ (\arrie)
\\
{\infer
{ }{\tyj{\x}{\x{:}\multiset{\sigma}}{\sigma}}
}
}
{\tyj{(\l \z.\s)\x}{\Gamma+\x{:}\multiset{\sigma}}{\tau}}
$$
\qed
\begin{lem}\label{lemma:HeandHew}
$(\Gamma, \sigma)$ is inhabited in $\Mu_e$ if and only if $(\Gamma, \sigma)$  is inhabited in 
$\Mu_{e,w}$.
\end{lem}
\proof
The {\it only if} part is trivial since any type derivation in $\Mu_e$ is also
a derivation  in  $\Mu_{e,w}$. Conversely, it is easy to show 
that any derivation $\Pi$ in  $\Mu_{e,w}$ may be mimicked in $\Su$, by induction 
on the size of $\Pi$: if $\Pi$  is an axiom,  then Lemma~\ref{lemma:H_w_increasing}  allows to conclude, otherwise the conclusion is by induction on the subderivations of the premises of $\Pi$'s last rule. \qed

Systems $\Mu_e$ and $\Mu_{e,w}$ inhabit exactly the same types, but $\Mu_{e,w}$ enjoys
subject reduction (Theorem~\ref{prop:ew}), and hence allow for searching inhabitants in normal
form.

\subsection{Disallowing untyped actual parameters }

In Section~\ref{section:fa} we have seen how strong normalization may be 
characterized by disallowing the empty (multi-)type.
This choice leads to the system  $\Mu_{e,w}$, in which a type is guessed 
arbitrarily whenever a term is $\lambda$-abstracted with respect to a fresh variable (\ie\ one which is not in the domain of the context). This corresponds to 
choosing an arbitrary type for the {\it formal} parameter of an erasing function. 
Dually, it is possible 
to characterize  strong normalization without disallowing the empty type, 
by just guessing an arbitrary type
for the {\it actual} parameters of all the  erasing functions.
The resulting system, called  $\Su$, is presented in Figure~\ref{fig:SN}, 
where types are defined as in Definition~\ref{def::lambdasys}. 

Along with system $\Mu_e$, system $\Su$ does not enjoy subject reduction, 
and this is certainly a major weakness. Still, it is a very simple system  characterising
strong normalisation, which has been considered for example in~\cite{BKV17,KV17}.

\begin{figure}[!ht]
\begin{center}
$\begin{array}{c}
\infer{\mbox{}}{\x:\multiset{\rho} \der \x:\rho}\ (\axvar)
\quad \quad 
\infer
{\Gam\der \s:\tau }
{\Gam \sm  \x \der \l \x. \s: \Gamma(\x)\arrow \tau}\ (\arri)\\ \\
\infer
{\Gam\der \s: \A  \arrow\tau \sep \Delta \der \uu:\A \sep \A \neq \ems  }
{\Gam + \Delta \der \s\uu: \tau}\ (\arrene) 
\hspace{1cm}
\infer
{\Gam\der \s: \ems  \arrow\tau \sep \Delta \der \uu: \multiset{\sig} }
{\Gam + \Delta \der \s\uu: \tau}\ (\arree)\\ \\
\infer{(\Delta_{i} \der \s:\sig_i)_{\iI}} {+_{\iI} \Delta_{i} \der \s:\multiset{\sigma_i}_{\iI}}(\many)
\end{array}$
\caption{The type assignment system $\Su$} \label{fig:SN}
\end{center}
\end{figure}

\ignore{As in system $\Mu_{e}$, the use of the choice operator in rule $(\arre)$
naturally gives the following two possible readings, where $\sig$ is an arbitrary type:
\[
\infer
{\Gam\der \s: \A  \arrow\tau \sep \Delta \der \uu:\A \sep \A \neq \ems  }
{\Gam + \Delta \der \s\uu: \tau}\ (\arrene) 
\hspace{1cm}
\infer
{\Gam\der \s: \ems  \arrow\tau \sep \Delta \der \uu: \multiset{\sig} }
{\Gam + \Delta \der \s\uu: \tau}\ (\arree) \]
}

Concerning the failure of subject reduction in system  $\Su$, the same counterexample given for system $\Mu_e$ applies:
$ \x:\multiset{\sigma}\ders (\l \y. \iid)\x:\multiset{\alpha} \arrow \alpha$, while
$ \x:\multiset{\sigma} \not\ders \iid:\multiset{\alpha} \arrow \alpha$. 

Notice that, if in the conclusion of rule $(\arree)$ we forget the context $\Delta$, then the system does
not characterize strong normalization anymore. In fact, 
starting from $\x:\mult{\emul \arrow \sigma}\ders \x:\emul \arrow \sigma $, we would
obtain  $\x:\multiset{\emul \arrow \sigma}\ders \x(\y\dup ):\sigma$ by  rule $(\mathtt{\arrow E_{\emul}})$, then 
 $\x:\multiset{\emul \arrow \sigma}\ders \l
\y.\x(\y\dup ): \emul \arrow \sigma$ by  $(\mathtt{\arrow I})$ , and finally
$\x:\multiset{\emul \arrow \sigma}\ders (\l\y.\x(\y\dup ))
  \dup: \sigma$, by $(\mathtt{\arrow E_{\emul}})$. The subject of the last judgement 
reduces to $\x(\dup\dup)$, which diverges.

In~\cite{BKV17} it is shown that the $\Su$-typable terms are exactly
the strongly normalizing ones, and the decidability of the inhabitation problem for 
$\Su$ is left open. We are going to show
that the inhabitation problem for $\Su$ is decidable, by reducing it
to the inhabitation problem for the system $\Su_{w}$ presented in
Figure~\ref{fig:SN,W}. System $\Su_{w}$ is obtained from $\Su$ by adding 
weakening to the axiom  so that, again, the usual weakening rule becomes admissible.

\begin{figure}[!ht]
\begin{center}
$\begin{array}{c}
\infer{1 \leq i \leq n}{\Gamma, \x:\multiset{\rho_1,...,\rho_n} \der \x:\rho_i }\ (\axvarw)
\quad \quad 
\infer
{\Gam\der \s:\tau }
{\Gam \sm  \x \der \l \x. \s: \Gamma(\x)\arrow \tau}\ (\arri)\\ \\
\infer
{\Gam\der \s: \A  \arrow\tau \sep \Delta \der \uu:\A \sep \A \neq \ems  }
{\Gam + \Delta \der \s\uu: \tau}\ (\arrene)  
\hspace{1cm}
\infer
{\Gam\der \s: \ems  \arrow\tau \sep \Delta \der \uu: \multiset{\sig} }
{\Gam + \Delta \der \s\uu: \tau}\ (\arree) \\ \\
\infer{(\Delta_{i} \der \s:\sig_i)_{\iI}} {+_{\iI} \Delta_{i} \der \s:\multiset{\sigma_i}_{\iI}}(\many)
\end{array}$
\caption{The type assignment system $\Su_w$}\label{fig:SN,W}
\end{center}
\end{figure}

 Before introducing the inhabitation algorithm for $\Su_{w}$,
  let us present the reduction of the inhabitation problem for system  $\Su$
  to the one for system $\Su_{w}$. The argument is the same as in the case
  $\Mu_e$ versus $\Mu_{e,w}$, treated in Section~\ref{section:sn1}.

\ignore{As before, rule $(\arre)$ can be split into two rules $(\arrene)$ and $(\arree)$. }
\begin{lem}\label{lemma:S_increasing}    Let $(\Gamma,\tau)$ be an instance of the inhabitation problem
having a solution in system $\Su$. Then 
 $(\Gamma+\x:\mult{\sigma},\tau)$ has also a solution in  $\Su$,
 for each variable $\x$ and type $\sigma$.
\end{lem}
\proof

A suitable type derivation is the following:
$$
\infer
{\infer
{\infer
{\vdots}{\tyj{\s}{\Gamma}{\tau}}
}
{\tyj{\l \z.\s}{\Gamma}{\emm\to\tau}}
\\
{\infer
{ }{\tyj{\x}{\x{:}\mult{\sigma}}{\sigma}}
}
}
{\tyj{(\l \z.\s)\x}{\Gamma+\x{:}\mult{\sigma}}{\tau}}
$$

\noindent $\z$ being a fresh variable.
\qed

\begin{lem}\label{lemma:SandSw}
$(\Gamma, \sigma)$ is inhabited in $\Su$ if and only if $(\Gamma, \sigma)$  is inhabited in 
$\Su_w$.
\end{lem}
\proof
The {\it only if} part is trivial since any type derivation in $\Su$ is also
a derivation  in  $\Su_w$. Conversely, it is easy to show 
that any derivation $\Pi$ in  $\Su_w$ may be mimicked in $\Su$, by induction 
on the size of $\Pi$: if $\Pi$  is an axiom,  then Lemma~\ref{lemma:S_increasing}  allows to conclude, otherwise the conclusion is by induction on the subderivations of the premises of $\Pi$'s last rule.
\qed

Systems $\Su$ and $\Su_w$ inhabit exactly the same types, but $\Su_w$ enjoys
subject reduction, and hence allow for searching inhabitants in normal
form.

\begin{prop}
The type system $\Su_w$ enjoys subject reduction.
\end{prop}
\proof
In order to show subject reduction, we first need a substitution lemma
that can be stated as follows. Let $\Gamma, \x:\A \der \s:\sigma$
and $\Delta \der \uu: \A$, then 
$\Gamma + \Delta \vdash \s\isubs{\uu}{\x}:\sigma$. 
The proof of this  lemma is by straightforward induction on $\Pi$.
It must be noticed that this proof uses the
admissible rule $(\mathtt{w})$. In fact, in case $\x \not \in \fv{\s}$, $\s\isubs{\uu}{\x}=\s$,
and in the resulting derivation $\Gamma + \Delta \vdash \s:\sigma$ all the premises in $\Delta$ 
need to be introduced by weakening. 

Now, in order to show subject reduction, we need to prove that
$\Pi \dem  \Gamma \der \s:\sigma$ and $\s \Rew{\beta} \s'$
implies there  exists a derivation $\Pi' \dem  \Gamma  \der \s':\sigma$.
The proof is by induction on the context $\ccontext$ such that $\s=\ccontext[(\l \x.\uu)\vv]$ and $\s' =\ccontext[\uu\isubs{\vv}{\x}]$.
The case $\ccontext=\square$ comes directly from the substitution lemma, the induction cases are straightforward.
\qed

Moreover, the type system $\Su_w$ characterizes strong-normalization.
\begin{prop}
\label{p:Sw-typ-sn}
Let $\s$ be a $\lambda$-term. Then $\Gam \der \s:\sigma$ iff $\s$ is strongly-normalizing. 
\end{prop}

\proof
Exactly the same reasoning used in Theorem~\ref{prop:ew}.
\qed

\subsubsection{Inhabitation for $\Su_w$}

In system $S_w$, types have in general infinitely many different inhabitants in normal
form. For instance, the problem
$(\x: \multiset{\emm\to\sigma},\sigma)$ admits all the solutions of the form
$\x\s$, where $\s$ is closed and normal. In order to be complete in
the sense of our previous Theorem~\ref{t:soundness-completeness-for-H}, the inhabitation
algorithm for $S_w$ should produce an arbitrary normal form anytime
the argument of an erasing function has to be constructed. Instead, we
decide to treat all such case uniformly, by always constructing the same
fake argument, namely the identity.

\begin{defi}
A type derivation $\Pi$ in  $\Su_w$ is {\it standard} if all instances of 
 $(\arree)$ in $\Pi$ have the shape

$$
\infer
{\Gam\der \s: \emul \arrow\rho \sep 
\Del \der \iid: \multiset{\alpha}\arrow\alpha }
{\Gam +\Del\der \s \iid: \rho}\
$$

We write $\Pi_{st}$ is $\Pi$ is a standard derivation.
\end{defi}

\begin{lem}
If $(\Gamma,\sigma)$ is inhabited in $\Su_w$, then there exists a standard derivation  $\Pi \triangleright {\Gam \der_{\Su_w} \s: \sigma} $
\end{lem}
\proof
Let  $\Pi' \triangleright {\Gam \der_{\Su_w}  \s': \sigma}$ be any derivation. Replacing the non-standard instances of  $(\arree)$ in 
$\Pi'$ of the form 

$$
\infer
{ \Gam\der \s: \emul \arrow\rho
 \sep 
 \Delta\der \uu:\sigma }
{\Gam +\Delta\derw \s\uu: \rho}\ (\arree)
$$
by 
$$
\infer
{
{\Gam \der \s: \emul \arrow\rho}
 \sep 
\infer{\infer{\ }{\Del+\x:\multiset{\alpha}\der \x:\alpha}\ (\axvar)}
{\Del \der \iid:\multiset{\alpha}\arrow \alpha }\ (\arri) }
{\Gam +\Delta\der \s \iid: \rho}\ (\arree)
$$
where $\x$ is fresh,
we obtain a standard derivation  for an inhabitant of $(\Gamma,\sigma)$.
\qed

The previous Lemma shows that, in order to decide whether a type is inhabited in  $\Su_w$, it is sufficient to look 
for standard derivations. This is what  the inhabitation algorithm  given in Figure~\ref{fig:inhab_sw} does.
The rule  $(\Head_{>0})$ splits in two: the usual one, used now for the case $\A \neq \emm$, 
 and the new rule 
$(\Head^{\emul}_{>0})$ introducing subterms of the form $\iid$ to be
taken as actual parameters 
of erasing functions. 
The soundness of the inhabitation algorithm holds in general, as for the other systems,
with the aside that the obtained derivation is standard. On the other hand, completeness holds in a relativized form:
only the normal forms of the subjects of standard derivations are reconstructed by the algorithm.
In order to establish completeness we will make use of the following
stability property:

\begin{lem}[Stability of Standard Derivations]
\label{l:stability}
If $\Pi \dem \Gam \der \s:\sig$ is a standard derivation and
$\s \Rew{\beta} \s'$, then not only there exists $\Pi' \dem \Gam \der \s':\sig$,
but $\Pi'$ is also a standard derivation. 
 \end{lem}

\begin{figure}[!ht]

\[ \infer{\s \real \K(\Gam +  \x:\A,   \tau)  \sep \sep 
   \x \notin \dom{\Gam} }
{ \l \x. \s \real  \K(\Gam  ,  \A \arrow \tau)}\ (\Abs)
\]

\[ 
   \infer{ (\s \real  \K( \Gam_i,  \sig_i))_{\iI} }
         {\s \real   \KI(+_{ \iI} \Gam_i,  \multiset{\sig_i}_{\iI})}\ (\Union) 
         \]
\[ \infer{
 \Gam = \Gam_{1}+\Gam_{2} \sep \s  \real  \LH{\x: \multiset{\A_1 \arrow \ldots \A_n\arrow 
\D \arrow \tau}}{(\Gam_{1}, \D \arrow \tau)} \sep \uu\real  \KI(\Gam_{2}, \D) \sep \D\not= \emul \sep n\geq 0
}
 { \s \uu  \real  \LH{\x: \multiset{\A_1 \arrow \ldots \A_n\arrow \D \arrow \tau}}{(\Gam, \tau)}}\ (\Head_{>0}) 
         \]
\[ \infer{
\s  \real  \LH{\x: \multiset{\A_1 \arrow \ldots \A_n\arrow 
\emul \arrow \tau}}{(\Gam_{1}, \emul \arrow \tau)}  \sep n\geq 0
}
 { \s\iid   \real  \LH{\x: \multiset{\A_1 \arrow \ldots \A_n\arrow \emul \arrow \tau}}{(\Gam, \tau)}}\ (\Head^{\emul}_{>0}) 
         \]        
\[ \infer{ }
         { \x  \real  \LH{\x: \multiset{ \tau}}{(\Gam, \tau)}}\ (\Head_{0}) 
         \]
         
         \[
     \infer{\s \real \LH{\x:\multiset{\A_1 \arrow \ldots \A_n \arrow \tau}}{(\Gam,\tau)}}
           {\s \real \K(\Gam + \x:\multiset{\A_1 \arrow \ldots \A_n \arrow \tau},\tau)}  \ (\StartHead) 
         \]

\caption{The inhabitation algorithm for $\Su_w$}
\label{fig:inhab_sw}

\end{figure}

%
%
%
%
%
%
%
%
%
%
%
%

As expected, the inhabitation algorithm terminates, a property that can be
shown exactly as in Lemma~\ref{l:termination}.
\begin{lem}[Termination]
\label{l:termination-Sw}
The inhabitation algorithm for system $\Su_w$ terminates. 
\end{lem} 

We now change the interpretation of the sets involved in the algorithm to:\\ 
 $\interp{\K(\Gam, \sig)} = \set{\s \mid \exists \Pi_{st}.\ 
\Pi_{st} \dem  \Gam \der \s:\sig   \mbox{ and }  \s    \mbox{ in normal form}    }$. \\
 $\interp{\KI(\Gam, \A)} = \set{\s \mid \exists  \Pi_{st}.\ 
\Pi_{st} \dem  \Gam \der \s:\ \mbox{ and }  \s    \mbox{ in normal form}}$. \\
$\interp{\Hn \x {\A_1\arrow\ldots\arrow A_n\arrow} \tau \Gam} = \set{ \x \s_1\ldots \s_n \mid 
\exists \Pi_{st}.\  \Pi_{st}  \dem  \Gam + \x:\mult{\A_1\arrow\ldots\arrow \A_n\arrow \tau}  \der \x\s_1...\s_n :\tau  \mbox{, }  
 \Gam=+_{i=1,\ldots,n}\Gam_i \mbox{, }  \Gam_i\der \s_i:\A_i \mbox{ and }  \s_i    \mbox{ in normal form}  }$
 

\begin{thm} [Soundness and Completeness for $\Su_w$]
  \label{t:soundness-completeness_for_Sw}
  \mbox{}
\begin{enumerate} 
\item \label{t:soundness-Sw}
If $\s\real  \K(\Gam, \sig)$  then $\s \in \interp{\K(\Gam, \sig)}$. 

\item \label{t:completeness-Sw}
If $\s \in \interp{\K(\Gam, \sig)}$, then  $\s\real \K(\Gam, \sig)$.
\end{enumerate}
\end{thm}
\proof
\begin{enumerate}
\item We  proceed analogously to the proof of Lemma~\ref{Lem:main}, proving the following statements by induction on the rules in Figure~\ref{fig:inhab_sw}, from which soundness follows directly.
\begin{enumerate}[label=\alph*)]
\item $\s\real \K(\Gam, \sig) \Rightarrow
\s \in \interp{\K(\Gam, \sig)}$.
\item $\s\real\KI(\Gamma,\A) \Rightarrow
\s \in \interp{\K(\Gam, \A)}$.
\item $\s \real \Hn \x {\A_1\arrow\ldots\arrow A_n\arrow} \tau \Gam \Rightarrow \s\in \interp{\Hn \x {\A_1\arrow\ldots\arrow A_n\arrow} \tau \Gam} $.
%
\end{enumerate}

The cases of rules $(\Abs)$ and $(\Union)$ are similar to the analogous cases in the proof of Lemma~\ref{Lem:main}, modulo considering normal forms instead than approximants.
There are three cases ending in the judgement
$\s \real \Hn \x {\A_1\arrow\ldots\arrow A_n\arrow} \tau \Gam$: 
\begin{itemize}
\item If $\x\real  \Hn \x {} \tau \Gam  $ comes from $(\Head_{0}) $, then 
$\Pi_{st} \dem \Gam+ \x:[\tau]  \vdash \x :\tau$ is obtained by rule $(\axvar_w)$. 

\item If  $\s\uu \real  \Hn \x {\A_1\arrow\ldots\arrow A_n\arrow} \tau {\Gam+\Delta}$
comes from rule $(\Head_{>0})$, where $\A_{n}\not = \emul$, 
$\s \real  \Hn \x {\A_1\arrow\ldots\arrow \A_{n-1}\arrow} {\A_n\arrow \tau} \Gam$ and 
$\uu\real \KI(\Delta, \A_n)$, then:

\begin{itemize}
\item
by the \ih (c), $\s=\x\s_1...\s_{n-1}$ and there exists $\Pi^{1}_{st}\dem \Gam + \x:[\A_1\arrow\ldots\arrow A_n\arrow\tau] \der \s : \A_n\arrow \tau$.

\item
by the \ih (b) there exists $\Pi^{2}_{st}\dem \Delta \der \uu:\A_n$.
\end{itemize}
So, using Rule  $(\arrene)$, we obtain a proof $\Pi_{st}\dem \Gam+\Delta + \x:[\A_1\arrow\ldots\arrow A_n\arrow\tau] \der \x\s_1...\s_{n-1}\uu:\tau$, and we are done.

\item If $\s\iid \real  \Hn \x {\A_1\arrow\ldots\arrow \A_{n}\arrow\emul\arrow} \tau {\Gam}$
comes from rule $(\Head^{\emm}_{>0})$, where the premise is  
$\s \real  \Hn \x {\A_1\arrow\ldots\arrow \A_{n}\arrow} {\emul\arrow \tau} \Gam$, then 
$\s= \x\s_{1}\ldots \s_{n-1}$ and by the \ih (c) 
there exists $\Pi_{st}$ such that $\Pi_{st} \dem \Gam + \x:[\A_1\arrow\ldots\arrow \A_{n}\arrow\emul\arrow \tau] \der \s: \emul\arrow \tau$. Then the proof follows by rule $(\arree)$.
\end{itemize}

\item Let $\Pi_{st} \dem \Gam \der \s:\sig$ ($\Pi_{st} \dem \Gam \der \s:\A$),
$\s$ being a normal form. 
   We prove  the following statements by
  induction on the typing derivations:

\begin{enumerate}[label=\alph*)]
\item $\s \in \interp{\K(\Gam, \sig)} \Rightarrow
\s\real \K(\Gam, \sig)$.
\item $\s \in \interp{\K(\Gam, \A)} \Rightarrow
\s\real\KI(\Gamma,\A)$.
\item $\s\in \interp{\Hn \x {\A_1\arrow\ldots\arrow A_n\arrow} \tau \Gam} \Rightarrow \s \real \Hn \x {\A_1\arrow\ldots\arrow A_n\arrow} \tau \Gam $.
\end{enumerate}

\begin{itemize}
\item Case $(\axvar)$. Then there is a standard type derivation of the following form: 
$$\infer{\mbox{}}{\Pi\dem \Del + \x:\multiset{\rho_1, \ldots, \rho_n} \der \x:\rho_i}\ (\axvar_w)$$
In this case $\Ap(\Pi)=\x$ belongs to 
$\interp{\K(\Del + \x:\multiset{\rho_1, \ldots, \rho_n},\rho_i)}$ and 
to $\interp{ \Hnp \x {\rho_i}  { \Del +  \x:\multiset{\rho_1, \ldots, \rho_{i-1}, \rho_{i+1}, \ldots, \rho_n}}  {\rho_i} }$,
so we need to show a) and c) respectively.

By rule $(\Head_{0})$ we have $\x\real \Hnp \x {\rho_i}  { \Del +  \x:\multiset{\rho_1, \ldots, \rho_{i-1}, \rho_{i+1}, \ldots, \rho_n}}  {\rho_i} $, then
by rule $(\StartHead)$ we also conclude
$\x\real\K(\Del + \x:\multiset{\rho_1, \ldots, \rho_n} ,\rho_i)$.

\item  Case $(\arri)$.  Then there is a standard type derivation of the following form: 

$$\infer
{\Pi'_{st} \dem\Gam\der \uu:\tau }
{\Pi_{st} \dem\Gam \sm  \x \der \l \x. \uu: \Gamma(\x)\arrow \tau}\ (\arri)$$
 
where also $\uu$ is in normal form. 
In this case,  $\l \x. \uu \in \interp{\K(\Gam \sm  \x, \Gamma(\x)\arrow \tau)}$
(and thus $\uu\in \interp{\K(\Gam, \tau)}$), 
so that we need to show a). 
By the \ih (a)  we have that $\uu\real \K(\Gam, \tau)$, and we conclude that  $\l \x. \uu \real \K(\Gam \sm  \x, \Gamma(\x)\arrow \tau))$
by rule  $(\Abs)$.

\item Case $(\arre)$. Then  $\s=\x\uu_{1}...\uu_{n}$ for some $n>0$, where 
 the $\uu_{i}$'s are in turn normal forms ($1\leq i \leq n, n\geq 0$).
There is a multi type $\A_n$ such that the standard typing
derivation $\Pi$ has the following form:
$$\begin{prooftree}
\Pi^{1}_{st} \dem \Gam_{1} \vdash \x\uu_{1}...\uu_{n-1}:\A_n \arrow \tau \sep
\Pi^{2}_{st} \dem \Gam_{2}\der \uu_{n}:\A_n
\justifies{\Gam \vdash \x\uu_{1}...\uu_{n}: \tau}
\using{(\arre)}
\end{prooftree}$$
where $\Gam = \Gam_1 + \Gam_2$. We analyse two cases:
\begin{itemize}
\item $\s \in \interp{ \Hnp \x {T_\tau} {\Gam'} \tau    }$,
where $\Gam = \Gam' + \x:\multiset{T_\tau}$. 
Then $\Gam_1 = \Gam'_1 + \x:\multiset{T_\tau}$, and
$\Gam' =  \Gam'_1 + \Gam_2$.  By the \ih (c),
  $\x\uu_{1}...\uu_{n-1} \real \Hnp \x {T_\tau }  {\Gam'_{1}} {\A_n\arrow \tau}$.

If $\A_n=\emul$, then $\Pi$  standard implies $\uu_{n}=\iid$, and the result 
then follows by rule $(\Head^{\emul}_{>0})$.

If $\A_n\neq \emul$, then  by the  \ih (b) 
  $\uu_{n}\real\KI(\Gamma_2,\A_n)$, so we get $\s \real \Hnp \x {T_\tau}  {\Gam'} {\tau}$  by rule
  $(\Head_{>0})$.
\item  $\s \in \interp{ \K(\Gam, \tau)}$. Then 
$\s \in \interp{ \Hnp \x {T_\tau} {\Gam'} \tau     }$ 
for some type $T_\tau$
such that $\Gam = \Gam' + \x:\multiset{T_\tau}$ as remarked after Definition ~\ref{Def:sem}. 
Then we conclude $\s \real \Hnp \x {T_\tau} {\Gam'} \tau$  by the previous point and 
$\s \real \K(\Gam, \tau)$
by rule $(\StartHead)$. 
\end{itemize}

\item Case $(\many)$: Then  $\Pi  \dem \Gam \der \s:\A$ implies  we have a standard derivation of the following form:
$$\infer{(\Pi^i_{st}i\dem\Delta_{i} \der \s:\sig_i)_{\iI}} {\Pi_{st}\dem +_{\iI} \Delta_{i} \der \s:\multiset{\sig_{i}}_{\iI}}(\many)$$
with $\Gam = +_{\iI} \Delta_{i}$ and  $\A= \multiset{\sig_{i}}_{\iI}$. 
By the \ih (a) we have that for all $\iI$, $\s\real
\K(\Delta_i, \sigma_i)$, hence we conclude $\s \real \KI(+_{\iI}
\Delta_{i},\multiset{\sig_{i}}_{\iI} )$ by rule $(\Union)$. \qed
\end{itemize}
\end{enumerate}

\section{Conclusion}\label{sec:concl}
In this paper we have studied the inhabitation problem for some
intersection type assignment systems for the $\l$-calculus, where
intersection is considered modulo associativity and commutativity, but
not idempotency. We proved that the problem is decidable in all
the considered cases. There is a plethora of intersection type
assignment systems that can be classified with respect to their
semantic power, \ie\  the class of terms they characterize and their
logical aspects. Concerning the latter, we focus on the lack or the
presence of weakening and of the empty multitype; for the former, we
consider systems characterizing either the solvable or the strongly
normalizing terms. 

Figure~\ref{fig:prism} represents all the systems
we took into consideration, lying on six vertices of the unitary cube,
whose axes represent the following features (the feature being present
if the corresponding coordinate has value 1):
\begin{itemize}[leftmargin=\leftmargin]
\item [$x$-axis:] weakening,
\item [$y$-axis:] empty multitype,
\item [$z$-axis:] characterizing strong normalization.
\end{itemize}

The unoccupied  vertex $\bullet$ (resp. $\circ$) corresponds to a system
without  (resp. with) weakening and without the empty type, that 
does not characterize strong normalization. 
Since such systems do not seem pertinent, the cube of the three features reduces to a  prism.
We have shown that the inhabitation problem is decidable for all 
the systems of the prism: for those enjoying the subject reduction
property, \ie\ for $\Mu$, $\Mu_w$, $\Su_w$ and $\Mu_{e,w}$, this is done by showing the soundness and completeness of a suitable terminating algorithm. On the other hand, the inhabitation problem of
$\Mu_e$ reduces to that of $\Mu_{e,w}$, and  the one  of
$\Su$ reduces to that of $\Su_w$. Remark that the inhabitation problem of $\Mu$ does not reduces to that of $\Mu_w$,
since the latter inhabits stricly more typings than the former, \eg\ the typing $(\x:\mult{\sigma,\tau}, \sigma)$ is inhabited in   $\Mu_w$ but not in  $\Mu$.

The starting point is the type system $\Mu$, which characterizes the class of solvable terms, and whose inhabitation algorithm has been originally presented in~\cite{DBLP:conf/ifipTCS/BucciarelliKR14}. 
That algorithm turns out to be  remarkably stable with respect to addition or deletion of all the considered features. Showing its robustness is one of the point of this work.
Another remarkable fact is that seemingly hard inhabitation problems,
like those of $\Mu_e$ and $\Su_w$, become easily tractable by simply
adding weakening to the type system.

\begin{figure}[!ht]
$
\xymatrix@!0{
&&_y&&&&
\\
&&&&&&
\\
&& \Mu \ar@{-}[rr]\ar@{--}[dd]\ar@{--}[uu]
& & \Mu_w \ar@{--}[dd]&&
\\
&\Su \ar@{-}[ur]\ar@{-}[rr]\ar@{-}[dd]
& & \Su_w \ar@{-}[ur]\ar@{-}[dd]&&
\\
&& \bullet\ar@{--}[rr]
& & \circ\ar@{--}[rr]&&_x
\\
&\Mu_e\ar@{--}[dl]\ar@{--}[ur] \ar@{-}[rr]\ar@{--}[uuur] 
& & \Mu_{e,w}\ar@{--}[ur]\ar@{-}[uuur] &&
\\
_z&&&&&&
}
$
\caption{A prism of non-idempotent type assignment systems}
\label{fig:prism}
\end{figure}

\subsection*{Acknowledgements}\label{sec:concl}
We would like to thank the anonymous reviewers for their useful criticisms and suggestions.

\renewcommand{\em}{\it}

\end{document}